\documentclass[sigconf,balance=false]{acmart}

\usepackage{graphicx}
\usepackage{xcolor}
\usepackage{enumitem}
\usepackage{hyperref}
\usepackage{booktabs}
\usepackage{soul}
\usepackage{algorithm,algpseudocode}
\usepackage{xparse}
\usepackage{fontawesome5}

\makeatletter
\NewDocumentCommand{\LeftComment}{s m}{%
  \Statex \IfBooleanF{#1}{\hspace*{\ALG@thistlm}}\(\triangleright\) #2}
\makeatother

\newcommand{\olio}[0]{\textsc{Olio}}

\newcommand {\change}[1]{{\color{black}#1\normalfont}}

\usepackage{nccmath} 

\newenvironment{tight_itemize}{\begin{itemize} \itemsep -1pt}{\end{itemize}}

\newcommand{\pheading}[1]{\vspace{4px}\noindent\textbf{#1}}
\newcommand*{\escape}[1]{\texttt{\textbackslash#1}}

\AtBeginDocument{%
  \providecommand\BibTeX{{%
    \normalfont B\kern-0.5em{\scshape i\kern-0.25em b}\kern-0.8em\TeX}}}

\copyrightyear{2023}
\acmYear{2023}
\setcopyright{rightsretained}
\acmConference[UIST '23]{The 36th Annual ACM Symposium on User Interface
Software and Technology}{October 29-November 1, 2023}{San Francisco, CA,
USA}
\acmBooktitle{The 36th Annual ACM Symposium on User Interface Software and
Technology (UIST '23), October 29-November 1, 2023, San Francisco, CA, USA}
\acmDOI{10.1145/3586183.3606806}
\acmISBN{979-8-4007-0132-0/23/10}

\begin{document}

\title{\olio: A Semantic Search Interface for Data Repositories}

\author{Vidya Setlur}
\affiliation{%
  \institution{Tableau Research}
  \city{Palo Alto}
  \country{USA}}
\email{vsetlur@tableau.com}

\author{Andriy Kanyuka}
\affiliation{%
  \institution{Tableau Software}
  \city{Vancouver}
  \country{Canada}}
\email{akanyuka@tableau.com}

\author{Arjun Srinivasan}
\affiliation{%
  \institution{Tableau Research}
  \city{Seattle}
  \country{USA}}
\email{arjunsrinivasan@tableau.com}

\renewcommand{\shortauthors}{Setlur, et al.}

\begin{abstract}
Search and information retrieval systems are becoming more expressive in interpreting user queries beyond the traditional weighted bag-of-words model of document retrieval. For example, searching for a flight status or a game score returns a dynamically generated response along with supporting, pre-authored documents contextually relevant to the query. In this paper, we extend this hybrid search paradigm to data repositories that contain curated data sources and visualization content. We introduce a semantic search interface, \olio, that provides a hybrid set of results comprising both auto-generated visualization responses and pre-authored charts to blend analytical question-answering with content discovery search goals. We specifically explore three search scenarios - question-and-answering, exploratory search, and design search over data repositories. The interface also provides faceted search support for users to refine and filter the conventional best-first search results based on parameters such as author name, time, and chart type. A preliminary user evaluation of the system demonstrates that \olio's interface and the hybrid search paradigm collectively afford greater expressivity in how users discover insights and visualization content in data repositories.
\end{abstract}



\begin{teaserfigure}
  \includegraphics[width=\textwidth]{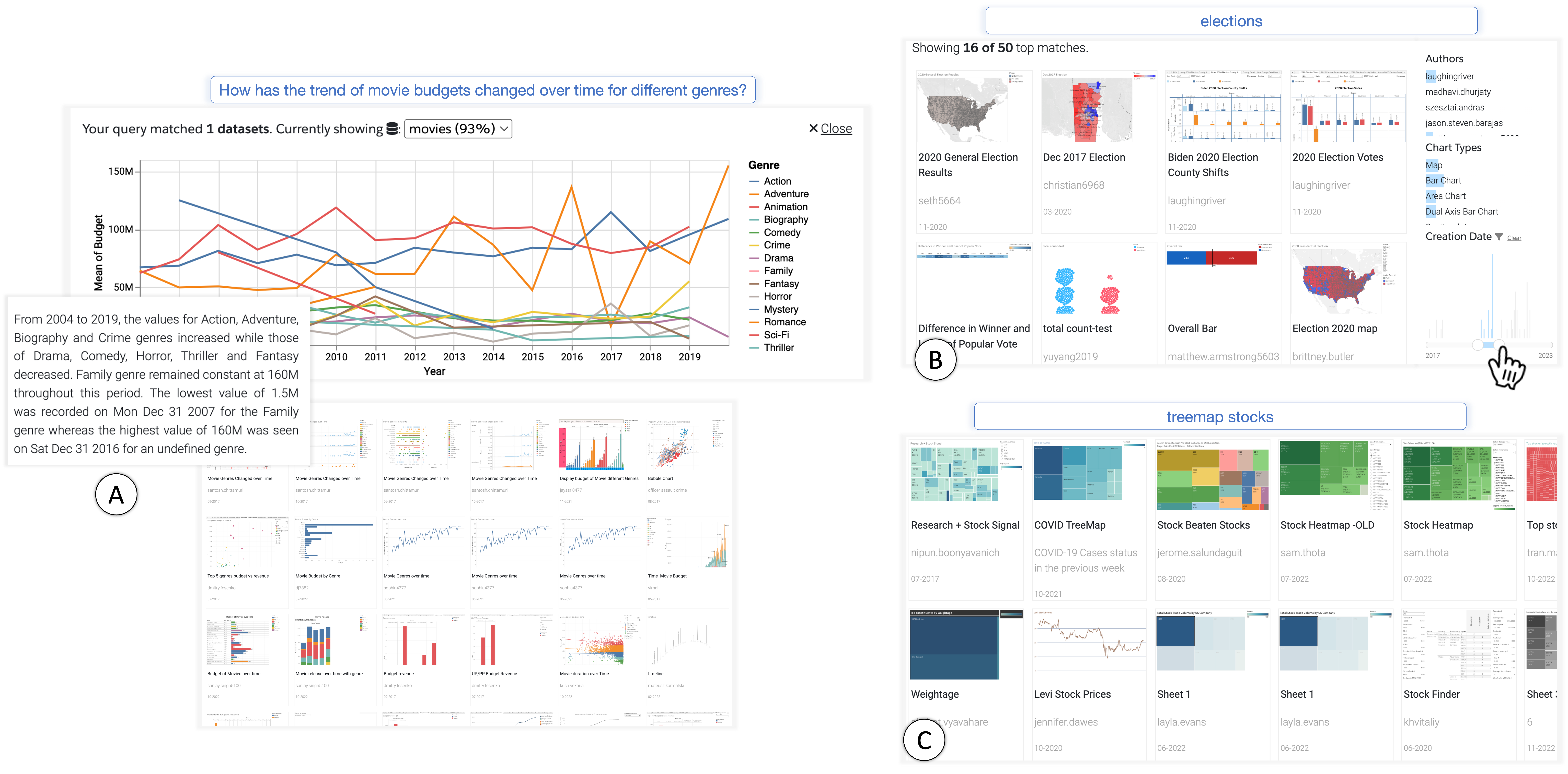}
  \caption{Examples of various semantic search scenarios supported in \olio{}. (A) \textit{Q\&A}. For an input query, \textit{``How has the trend of movie budgets changed over time for different genres?,''} \olio{} detects that it is a Q\&A search with an analytical intent, `trend.' A curated data source, `movies,' is the top-scored match to the query, and the system generates a multivariate line chart  response. A generated text summary describes the visualization as shown. Pre-authored visualization content is also displayed as thumbnails below the generated response as additional information. (B) \textit{Exploratory Search}. \olio{} identifies the input query, \textit{``elections,''} as a keyword search query and shows pre-authored visualizations with text content pertaining to `elections.' (C) \textit{Design Search}. The query, \textit{``treemap stocks''} is identified as a search of all content containing treemap visualizations pertaining to `stocks.' \olio{} returns a set of relevant pre-authored visualizations for the query and displays them as thumbnails. The thumbnails are linked to the actual visualizations if the user desires to continue with their analytical workflow.}
  \Description{}
  \label{fig:teaser}
\end{teaserfigure}


\maketitle

\section{Introduction}
User expectations of search interfaces are evolving. Search engines are increasingly expected to answer questions along with providing contextually relevant content that help address a searcher's goal~\cite{crook:2018}. Existing keyword-based search methods are mostly designed for content retrieval. Their main underlying drawback is limited support for structured query types that generally expect focused and specific responses. Natural language (NL) question \& answering (Q\&A) interfaces, on the other hand, support more fact-finding inquiry but do not support content or document discovery and retrieval. To bridge the gap between these two contrasting search paradigms, a hybrid approach called \emph{semantic search}~\cite{Guha2003SemanticS} applies user intent and the meaning (i.e., semantics) of words and phrases to determine the right content that might not be present immediately in the text (the keywords themselves) but is closely tied to what the searcher wants~\cite{bhagdev:2008}. The information retrieval technique goes beyond simple keyword matching by using information such as entity recognition, word disambiguation, and relationship extraction to interpret the searcher's intent in the queries. For example, keyword search can find documents with the query, ``French press,'' while queries such as ``How do I make quickly make strong coffee?'' or ``manual coffee brewing methods'' are better served by semantic search to produce targeted responses.

With an increase in the number of data repositories on the web, including structured data in the form of relational databases, files, and knowledge graphs, there is a plethora of information that supports the blend of generating responses to fact-finding questions with document retrieval~\cite{galhotra:2020}. Along similar lines, data repositories and visualization tools such as Observable~\cite{Observable}, Tableau Public~\cite{tableau2023public}, and Microsoft Power BI Partner Showcase~\cite{powerbi} host hundreds or thousands of visualizations representing a wide range of datasets, making them rich platforms for knowledge sharing and consumption. Search plays a pivotal role in these repositories, providing people the ability to winnow in on content they are interested in (e.g., charts on a specific topic, charts showing data trends and bespoke visualizations such as Sankey diagrams, or charts authored by a particular person). Current search systems tend to rely on document-retrieval techniques to provide relevant search results for a given query. However, the challenge with data repositories lies in the sparseness of searchable text within them; data sources and charts often have limited text information in the form of titles, captions, and textual data values, for example. There is a need to explore alternative ways to index and search for content based on this limited availability of textual information.

Another challenge is that current search features for data repositories offer limited expressivity in specifying search queries, restricting users to predominantly perform keyword search for content based on the visualizations’ titles and authors. In contrast, other contemporary search interfaces such as general web search, image and video search, and social networking sites enable users to find and discover content through a rich combination of textual content (e.g., keywords or topics covered in a website), visual features within the content (e.g., looking for images with a specific background color), dates (e.g., only viewing videos from the recent week), geographic locations (e.g., limiting search to certain zip codes or cities), and even different types of media (e.g., searching for similar images through features like reverse image search). 

Designing expressive search interfaces for data repositories requires gaining a deeper empirical understanding of people's search requirements, given the current limitations of these systems. For instance, what goals do people have in mind when using search in the context of data repositories? How do people formulate their search queries? Is text alone a sufficient modality for search? If not, what are complementary/alternative modalities to consider? What supporting metadata do people want to query for or use to filter the search results?

\pheading{Contributions.} To explore these research questions, we first conducted a set of formative user elicitation interviews with $14$ participants who regularly search for visualizations or are involved in the design of search interfaces within mainstream visualization tools and data repositories. Findings from the interviews identified search scenarios specific to content exploration for data repositories and motivated the design and implementation of \olio\footnote{The word \textit{Olio} is defined as `a miscellaneous collection', reflecting the hybrid mix of search content displayed in the interface~\cite{mw:olio}}, an interface that supports semantic search behavior by dynamically generating visualization responses and pre-authored visualizations for data repositories. Specifically, the interface implements three search scenarios on a semantic search framework: \emph{Q\&A search} by interpreting analytical intent over a set of curated data sources, \emph{exploratory search} using document-based information retrieval methods on existing indexed visualization content, and \emph{design search} by leveraging visualization metadata for the content (Figure~\ref{fig:teaser}). The interface also supports facet-driven browsing to prune the search results by author name, time range, and visualization type. Employing \olio~as a design probe, we conducted a qualitative study with $11$ participants to gain feedback on the implemented metadata and querying features, identify system design and implementation challenges, and better understand user behavior.

The study confirmed that the semantic search paradigm supports the different data repository search goals. We observed that the ability to perform both Q\&A and search for pre-authored content facilitated a fluid analytic search experience but raised new questions about user expectations from search systems and the style of user interaction.
Lastly, from our observational data and participant feedback, we highlight promising directions for future work on visualization search interfaces, including better support for creating curated data sources, the need for scaffolding \change{(i.e., support to help with the discoverability of features or functionality in a user interface~\cite{sneakpique})} and building trust in the system behavior and exploring additional search paradigms and modalities.

\section{Related Work}
Prior research relating to search systems in the context of visual analysis generally falls into three main categories: (1) semantic web search systems, (2) natural language interfaces (NLIs) for visual analysis, and (3) search interfaces for visualizations.

\subsection{Semantic Web Search Systems}
Semantic search was initiated as a document search technique to improve searching precision by understanding the purpose
of the search (i.e., intent) and the contextual significance of words as they appear in the searchable data space to generate more relevant results~\cite{Guha2003SemanticS}. Approaches to semantic web search can roughly be divided into those systems based on structured query languages~\cite{corby:2004,swoogle,Heflin2003SHOEAB,Kasneci2008NAGAHS,oren:2008,Thomas2007ONTOSEARCH2SO}, keyword-based approaches~\cite{falcons,Harth2007SWSEAB,Lei2006SemSearchAS,Tran2007OntologyBasedIO,zenz:2009}, where queries consist of lists of keywords, and natural-language-based approaches~\cite{cimiano:2008, Damljanovic2010NaturalLI,fernandez:2008,Lpez2005AquaLogAO,lopez:2006}. Our work is inspired by the body of semantic search techniques where we explore how we can support various search scenarios (i.e., Q\&A, exploratory, and design search) for exploring data repositories.

Early research focused on the problem of augmenting traditional text search with additional metadata using ontological techniques to increase recall and precision~\cite{moldovan:2000,Buscaldi2005AWQ,grubar:1993,ciri}. Our work explores semantic augmentation in the context of data repository search by considering additional metadata pertaining to attributes in curated data sources, such as synonyms and related concepts, as well as metadata that describes the pre-authored content, such as the visualization type, data attributes, and author name.

To support targeted Q\&A in semantic search, systems have explored ways for accurately detecting NL patterns and phrases that represent temporal intents such as "in the 20th century" or spatial intents such as "in Europe''~\cite{KOLOMIYETS20115412}. Typical approaches in this direction involve a combination of statistical techniques (syntactic parsing) and semantic operations to identify ontology concepts in the user's input. For instance,
QUERIX~\cite{kaufmann:2006} combines the Stanford CoreNLP parser with WordNet to recognize salient phrases from NL user queries~\cite{klein-manning-2003-accurate}. Other Q\&A  systems apply linguistic processing to the question, identifying named entities and other query-relevant phrases~\cite{Srihari1999InformationES,chu-carroll:2006,panto}. \olio~identifies a set of analytical intents (e.g., trends, location, groupings, aggregations, filters) in the queries for supporting Q\&A in a data-oriented semantic search context.

More recently, web search engines blend complementary search experiences of machine-generated results with pre-authored documents and web pages~\cite{ding:2005}. Search platforms~\cite{google,Bing} have made updates to their search algorithms that place greater emphasis on search queries, considering overall context and meaning over individual keywords. The algorithm employs form-based or `template' queries to answer questions at scale in real-time such as the weather, flight status, or the current score of a basketball game. The premise of our research is to explore a similar search paradigm, specifically in the context of data repositories, where we explore the interpretation of queries containing bespoke analytical intents in addition to keyword search. 

Traditional information retrieval methods rely on large amounts of searchable text content. However, multimedia repositories that include videos and images, have limited searchable text content. To this end, research in multimedia retrieval has explored metadata extraction techniques to improve the precision and recall of the search algorithms. Techniques include constructing bag-of-word image descriptors from the associated text in documents referring to other similar images~\cite{vaca-castano:2015}, analyzing visual features in images~\cite{visualrank}, parsing XML descriptors in MPEG video files~\cite{hammiche:2004}, and object and scene retrieval in videos~\cite{videogoogle}, to name a few. Our work addresses an analogous problem when searching data repositories, given the sparseness of searchable text content. We include additional semantics for both data sources and visualizations using ontological enrichment from external corpora, along with properties extracted from the XML properties in the visualizations.

\subsection{NLIs for Visual Analysis}  
NLIs for visual analysis specifically support dynamic Q\&A in the larger context of semantic search experience. Systems like DataTone~\cite{datatone} support analytical Q\&A, producing a chart according to that inference and then providing ambiguity widgets through which the user could adjust the system's default choice. Eviza~\cite{eviza} and Analyza~\cite{analyza} extend that premise through contextual inferencing. Evizeon~\cite{hoque2017applying} and Orko~\cite{orko} explore the notion of pragmatics in analytical conversation by using the knowledge of data attributes, values, and data-related expressions. 

Commercial visualization Q\&A systems~\cite{thoughtspot,ibmwatson,powerbi} have evolved over the years to better understand a user's analytical intent expressed in NL and provide reasonable visualization responses. The forms of inferring intent typically rely on explicitly named data attributes, values, and chart types in the user's input queries. Ask Data~\cite{setlur2019inferencing} handles various analytical expressions in NL form, such as grouping of attributes, aggregations, filters, and sorts. The system also handles impreciseness around vague numerical concepts such as `cheap' and `high' by inferring a range based on the underlying statistical properties of the data. 

However, these systems assume that the data source or dashboard is already preselected before interpreting the queries. Further, they tend to focus on a subset of semantic search (primarily Q\&A). Our work explores how analytical search intent can be interpreted to support the various flavors of search across multiple repositories of data sources and visualizations.

\subsection{Search interfaces for Visualizations}

Large-scale search platforms for visualizations have focused on experiences to help users reason and analyze data sets of interest. ManyEyes, a web-based service, combined public data sharing with interactive visualizations~\cite{viegas2007manyeyes}. Users could upload and visualize data on the web, facilitating the sharing
and discussion of visualizations. Morton et al.~\cite{morton2014public} used Tableau Public as a platform to analyze the use of online visual analysis systems and point out that there is a need for improvement of web-based visualization analytics systems to better support both search and content diversity of visualization designs. 

Past research also highlights a need for search tools and interfaces to be better integrated into users' authoring workflows. Battle et al. suggest new user experiences, such as design search where the visualization community could easily find D3 content based on chart types, visual style, and structure to help translate their ideas into often complex and bespoke visualizations~\cite{battle2021exploring}. Hoque and Agrawala~\cite{hoque2019searching} present a search engine for D3 visualizations collected from the web that allows queries based on their visual style and underlying structure. Their search engine indexes the marks and encoding, along with visual style and layout, to support the exploration of D3 charts with specific design characteristics. SightLine is a web portal that passively collects and organizes visualizations to explore the design space of visualizations on the web~\cite{sechler2017sightline}. By preserving the context of each visualization visit, the tool enables personal provenance through the discovery and exploration of trending visualizations, as well as a more targeted search by querying the metadata collected for each visualization. Along these lines, Observable has the provision for specifying search tags to restrict and combine search terms~\cite{Observable:searchtags}. The tags stem from metadata properties for these notebooks, including author, title, collection name, etc.~\cite{Observable:metadata}.

Building on prior research, our work recognizes the various scenarios for search in the context of data repositories and explores a semantic search user experience for supporting these scenarios within a \emph{unified} interface. \olio~serves as a research probe to explore the interpretation of search intent against data sources and visualizations by utilizing their underlying metadata. 

\section{Identifying Search Scenarios for Data Repositories}
\label{sec:formative}

To better understand the types of search tasks people would find useful when searching over data repositories, we conducted a series of interviews.
We sought to collect a broad perspective from users spanning different backgrounds (e.g., programmers vs. non-programmers) and roles (e.g., visualization designers, consultants, casual viewers, or consumers).
We recruited $14$ participants \change{(7 females, 7 males)}, including seven visualization designers or consultants, three product managers involved in the design of visualization repositories, and four software engineers and designers. Participants had working experience with visualization repositories for tools like Tableau (e.g., Tableau Public), Microsoft Power BI (e.g., Power BI Partner Showcase), D3 (e.g., D3's Observable Example Gallery), and general experience searching for visualizations on Google.

Interviews were conducted remotely and lasted 30-45 minutes. We asked participants about their backgrounds (e.g., their job descriptions, visualization repositories they use actively) and then asked them to share their experience, including the scenarios in which they search data or visualization repositories, current limitations, and areas for improvement in terms of the search experience, and metadata they find most relevant during visualization search.
We qualitatively analyzed the session transcripts and used an affinity diagramming approach to iteratively group similar comments (e.g., comments referring to searching for visualizations with a specific title or by an author, comments referring to using chart type as part of the search query).
We combined these groups under broader clusters of different scenarios search is used in as well as the most relevant search querying features.
Below, we summarize the key findings from our formative interviews in terms of the user goals and metadata features most relevant to \change{search in the context of data repositories containing both datasets and pre-authored charts}.

\subsection{Search Scenarios}

We identified three key user goals or scenarios for search in the context of data repositories.

\begin{itemize}[leftmargin=.15in]
    \item \textbf{Question \& Answering (Q\&A).}
    One common goal echoed by participants, particularly those who worked with organization-specific repositories hosting several data sources, was to leverage search to answer analytic questions.
    This goal is similar to information lookup~\cite{hearst2009search} in the broader web search context where user queries map to brief and discrete pieces of information (e.g., entities, dates, computed values).
    However, with data repositories, participants wanted to issue analytic questions (e.g., \textit{``What are sales trends across regions?,''} \textit{``highest covid cases by country''}) and get an appropriate response containing visualization and/or text generated from the available data sources.

    \vspace{.5em}
   \item \textbf{Exploratory Search.}
    In line with the notion of exploratory search in web search~\cite{marchionini2006exploratory}, participants wanted to leverage data repositories to learn about a topic through available charts and data.
    Examples of exploratory search queries include \textit{``NFL drafts,''} \textit{``USA covid trends,''} or \textit{``Fifa world cup.''}
    Such queries are typically open-ended and do not provide refined filtering criteria beyond the topic itself. For instance, one participant (a visualization consultant) referred to exploratory search as one of his prominent goals during the initial stages of customer interactions.
    He highlighted the example of searching for visualizations on \textit{``private equity dashboards"} on Tableau Public during his recent interaction with a client at an investment firm.
    Describing her use cases for search, another participant (a visualization designer) alluded to exploratory search as one of her frequent search goals, stating \textit{``I often use search to see a few examples of what people create and to hunt for data sources about a topic."}
    
    \vspace{.5em}
    \item \textbf{Design Search.}
    The ability to find visualizations based on design features (e.g., chart type, color) was another popular use case for search, especially among the seven participants who were designers/consultants or novice visualization authors.
    Design search query examples include \textit{``sunburst chart,"} \textit{``bar and line combination chart,"} or \textit{``map with icons."}
    Based on anecdotes shared by the participants, this type of search is typically performed when users are looking for learning resources (e.g., a novice D3 developer looking for examples of force-directed layouts created with D3, a Tableau user trying to create a bespoke visualization like a Sankey diagram) or trying to understand design practices and find inspiration for their own work (e.g., using searches like \textit{``maps with a dark background''} to find examples of charts with specific color constraints).
    
\end{itemize}

\noindent{}Note that these scenarios are neither exhaustive nor mutually exclusive.
For instance, \change{three participants mentioned ``targeted search'' as another scenario, where the intent was to retrieve a specific chart or dataset that the users knew existed in the repository.
However, we do not explicitly call this scenario out as it would inherently be supported by any search system that supports exploratory search (e.g., users can include specific and precise terms during exploratory search to retrieve the desired content). Furthermore,} 
queries like \textit{``sales by state and segment as a heatmap"} or \textit{``maps showing covid trends,"} combine Q\&A and design search, and design and exploratory search, respectively.
We also asked participants to rank the scenarios in terms of frequency/importance.
The responses, however, were fairly mixed and there was no single primary goal or a specific ordering of scenarios that stood out across participants.

Thus, rather than being a definitive and ordered set, the three scenarios listed above are primarily intended to serve as guidance for broad categories of user tasks to keep in mind when designing search systems for data repositories.

Besides understanding \textit{when} and \textit{why} people use search in data repositories (i.e., the above scenarios), to design and implement an effective search system, we also wanted to identify \textit{what} information people find most relevant while searching and browsing visualizations.
To this end, combining the participants' comments and search documentation for platforms like Observable~\cite{Observable:metadata} and Tableau Server~\cite{tableau_server_search}, we curated a list of the most prominent metadata fields that we focus on in our prototype.
These fields include the visualization title and description, the chart type (e.g., `bar chart,' `map,' `heatmap'), graphical encodings such as mark type, the visualization author, and the chart's creation date.

\subsection{Design Considerations}
Combining the feedback from the formative interviews with guidelines and findings from prior work on visualization search (e.g.,~\cite{viegas2007manyeyes,hoque2019searching,sechler2017sightline}), web and image search interfaces (e.g.,~\cite{hearst2009search,silverstein1999analysis,marchionini2006exploratory}), and NLIs for visual analysis (e.g.,~\cite{tory2019mean,shen2021towards,srinivasan2021collecting}).

\pheading{DC1. Support a unified experience that supports all three search scenarios.}
As we discussed the different search scenarios during the formative study, participants noted that they would ideally want the same interface and modality to perform the different tasks.
Thus, one consideration for us while building \olio{} was to design a seamless experience that supported a common input modality (NL) and blended Q\&A (a task commonly performed on data source collections) with exploratory and design search (tasks commonly performed with pre-authored visualization repositories).

\pheading{DC2. Support linguistic variations in queries.}
Both prior work on NLIs for visualization (e.g.,~\cite{tory2019mean,setlur2019inferencing,srinivasan2021collecting}) and web search (e.g.,~\cite{silverstein1999analysis,barr2008linguistic}) has shown that people use a variety of phrasings in search queries to accomplish the same goal.
Even during our interviews, participants used linguistically varied examples while discussing the same goal (e.g., ``\textit{What are sales trends across regions?}'' vs.~``\textit{sales by region over time}'').
Accommodating such user behavior, a second design consideration for \olio{} was that the system should support a variety of query formats - terse keywords as well as queries phrased as questions or sentence fragments, with an understanding of analytical intent relevant to data repositories in either case.

\pheading{DC3. Show textual responses and provide guidance for Q\&A queries.}
When discussing Q\&A scenarios, we asked participants about the types of visualizations they would expect for different queries.
During these conversations, in line with prior research on information lookup on the web~\cite{marchionini2006exploratory}, participants noted that besides charts, it may be valuable to ``\textit{provide a text response to a text query},'' suggesting the inclusion of complementary text along with a generated chart.
To this end, we noted that given a Q\&A query, \olio{} should not only select an appropriate data source and generate a chart but also text content that leverages the chart to help answer the input query.
Furthermore, since Q\&A queries can map to multiple data sources and users may not be aware of the available data source and fields, the system should guide users to ask questions (e.g., via query suggestions) and provide metadata information on the relevant data sources (e.g., available data fields and values to query).

\pheading{DC4. Provide visual summaries and filtering options for search results.}
One struggle that was echoed by several participants was that current visualization search systems do not provide an easy way to comprehend and sift through results beyond manual inspection.
To overcome this limitation with current systems, we noted that the system should provide visual summaries and support dynamic filtering~\cite{ahlberg1994visual} to help people get an overview, organize, and create meaningful facets of the visualization search results.

\section{\olio}

\olio~is designed as an interface that supports semantic search behavior by dynamically generating visualization responses and pre-authored visualizations from data repositories. Below, we describe \olio{'s} interface through a brief usage scenario and subsequently detail the key system components and implementation.

\subsection{Interface}

\begin{figure}[t!]
    \centering
    \includegraphics[width=\linewidth]{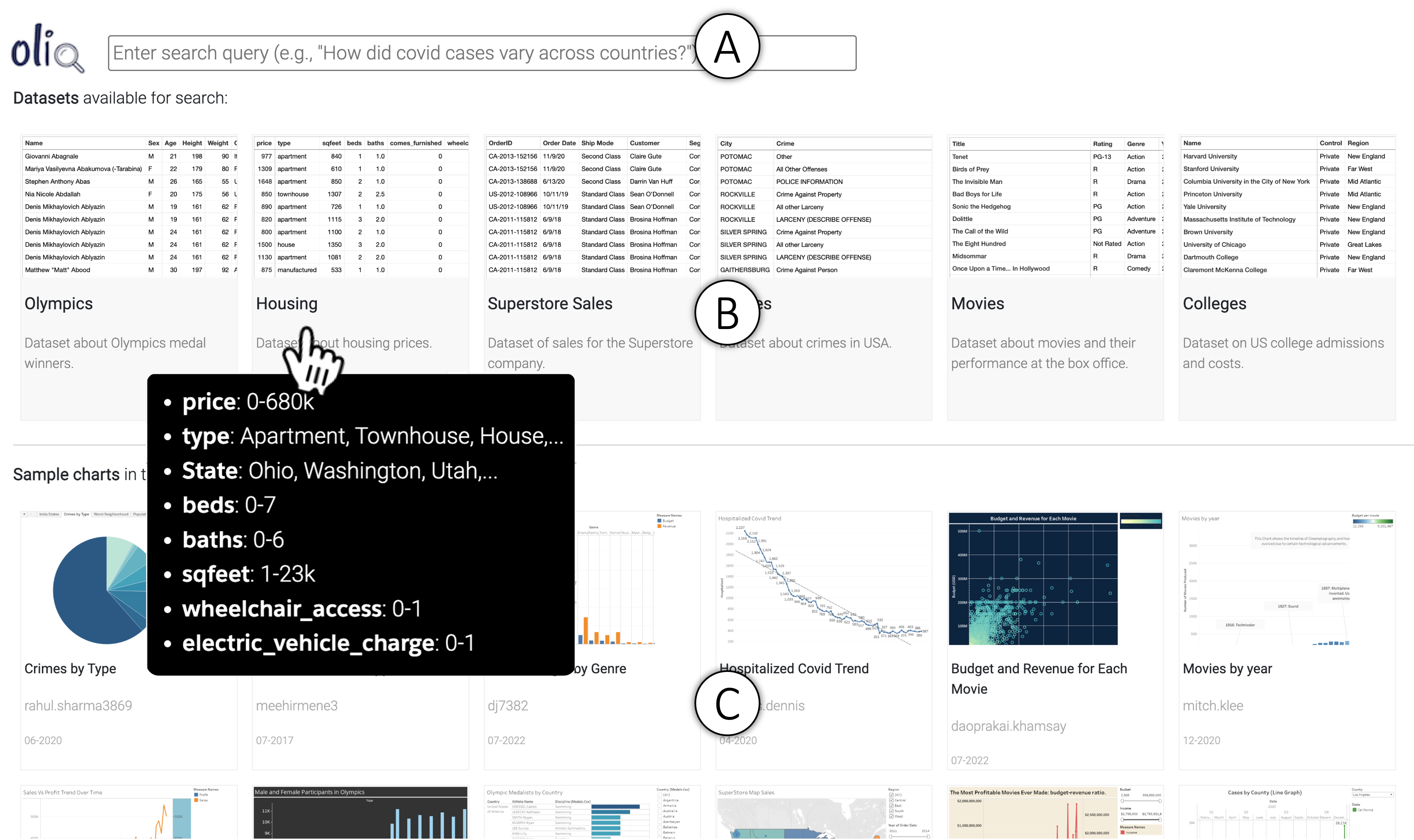}
    \caption{\olio{'s} landing screen.
    (A) A search input box with a placeholder query suggestion generated based on one of the available data sources.
    (B) Thumbnail previews of some available data sources. Here, hovering over the `Housing' data source shows a tooltip displaying metadata about the data source's attributes and values.
    (C) A sampling of pre-authored visualizations available for search.}
    \label{fig:start-screen}
\end{figure}

\begin{figure}[b!]
    \centering
    \includegraphics[width=\linewidth]{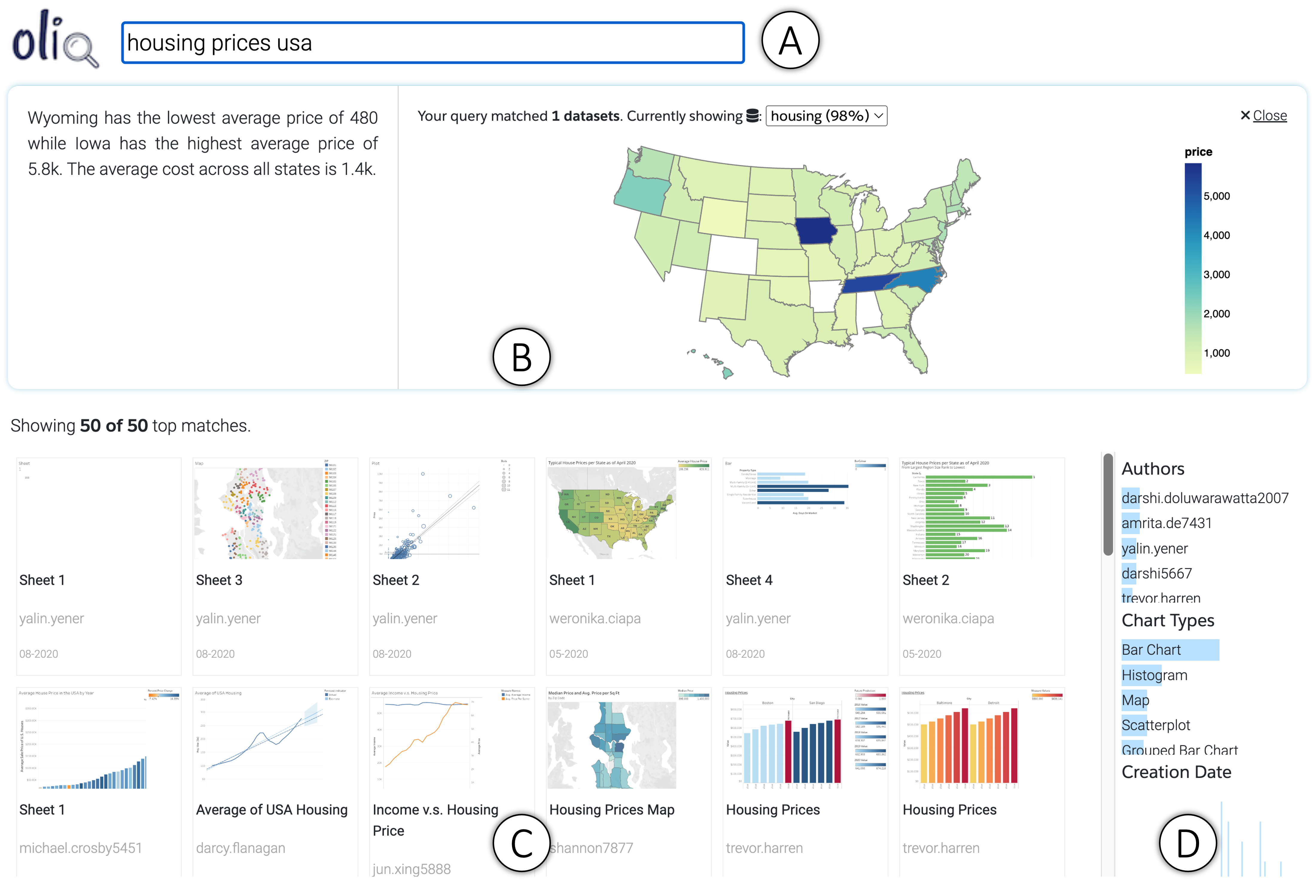}=
    \caption{The \olio{} interface.
    (A) Search input box.
    (B) Dynamically generated content, including a chart on the right and text highlighting the key takeaway messages from the chart on the left. Users can hover the mouse cursor over the {\small{\faDatabase}} icon to display a dataset summary tooltip similar to Figure~\ref{fig:start-screen}B.
    (C) Top 50 pre-authored visualizations that map to the input query.
    (D) Scented widgets that support dynamic filtering of the pre-authored content results.}
    \label{fig:interface}
\end{figure}

The interface initially shows a landing screen that displays a sampling of data sources available for Q\&A search (Figure~\ref{fig:start-screen}). A user can hover over a data source thumbnail and view its corresponding metadata information (\textbf{DC3}). The user then types a search query, \textit{``housing prices usa''} in the input text box (Figure~\ref{fig:interface}A). The system detects that token `usa' is a geographic location and searches for a relevant data source in its data repository. \olio~finds the housing data source to be a match, and a map is dynamically generated as a Q\&A response to the query (Figure~\ref{fig:interface}B). In addition, as part of exploratory search, the query tokens are used as keywords to match any pre-authored visualizations (\textbf{DC1}). A grid of thumbnails is displayed to serve as a preview to the user for browsing and exploration (Figure~\ref{fig:interface}C). Each thumbnail is hyperlinked to its corresponding visualization file that the user can choose to peruse in more detail or download to their local machine. The title, author name, and creation date of the visualization are displayed below each thumbnail to provide additional context. Scented widgets~\cite{willett2007scented} appear on the right side of the exploratory search panel to support faceted browsing of the pre-authored visualizations (Figure~\ref{fig:interface}D). The user can narrow down the search results by simultaneously applying one or more filters, namely, author name, visualization type, and the creation date (\textbf{DC4}).

\subsection{System Overview}
\begin{figure}[ht]
    \centering
    \includegraphics[width=\linewidth]{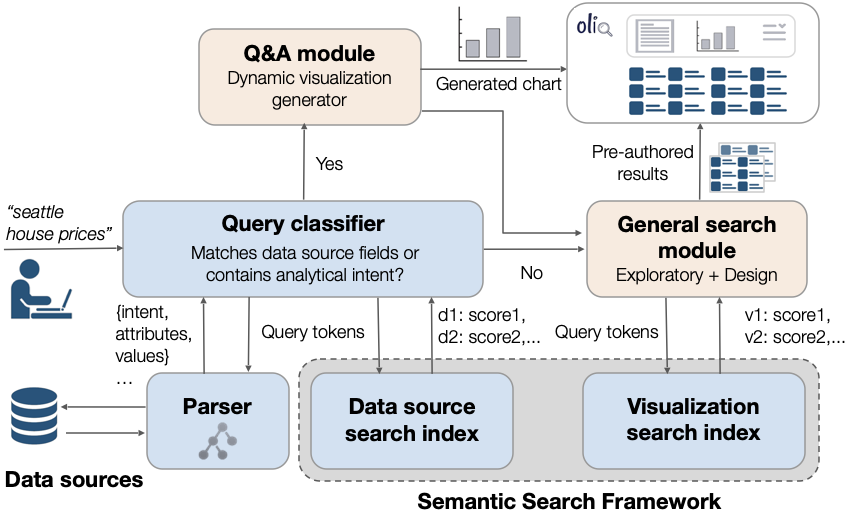}
    \caption{System architecture overview showing the various components: query classifier, parser, semantic search framework, Q\&A and general search modules. The query classifier first checks for the presence of tokens that refer to fields from the data sources and analytical intents from the parsed query. If present, dynamically generated visualizations from Q\&A search component are rendered, along with general search components (exploratory and design), returning pre-authored visualization results.}
    \label{fig:overview}
\end{figure}

\olio~is implemented as a web-based application using Python and a Flask backend connected to a Node.js frontend. We leverage Elasticsearch~\cite{elasticsearch}, an open-source Java search engine that is designed to be distributive, scalable, and with near real-time query execution performance. \change{As a result, \olio{} can scale to a large number of data repositories for indexing and search.} Figure~\ref{fig:overview} illustrates a high-level depiction of the system’s architecture, with the following main components: query classifier, parser, semantic search framework, Q\&A module, and the general search module.

A repository of curated data sources is included in the system for Q\&A search. The data sources could be any tabular CSV file, but for the purpose of this prototype, we include \change{eight} data sources across a variety of familiar topics such as sales~\cite{superstore,coffeesales}, sports~\cite{rgriffin_2018}, world events~\cite{covid-ca}, entertainment~\cite{bansal_2021}, and civic issues~\cite{us-crimes, housing}. \change{The datasources varied in the number of attributes as well as their cardinality, including 4-20 columns and $\sim$300-28,000 rows.}


\subsection{Data Repositories and Metadata}
\label{sec:metadata}
Unlike traditional document search, data sources and visualizations tend to be text-sparse, with limited searchable text content. Hence, \olio~augments the data repositories with additional metadata and semantics that helps the system's understanding and interpretation of the search queries. Specifically, attributes and values in the data sources are linked to ontological concepts, including synonyms (e.g., `film' and `movie')~\cite{thesaurus} and related terms (e.g., `theft,' `burglary,' and `crime')~\cite{word2vec}. The system includes a small hierarchy of hypernyms and hyponyms, from Wordnet~\cite{wordnet}, whose depth typically ranges up or down to two hierarchical levels (e.g., $[`beverage,' `drink'] \rightarrow [`espresso,' `cappuccino']$).
The metadata also includes data types (i.e., `text,' `date,' `Boolean,' `geospatial,' `temporal,' and ‘numeric’) and attribute semantics, such as currency type (e.g., United States Dollar). This information could also be inferred using existing data pattern matching techniques~\cite{pytheus,adelfio:2013,potterswheel}. The metadata also identifies attributes that are measures (i.e., attributes that can be measured, aggregated, or used for mathematical operations) and dimensions (i.e., fields that cannot be aggregated except as count). This final set of metadata information is then added to the semantic search framework.

The pre-authored content is a set of $75,000$ visualizations sourced from Tableau Public~\cite{tableau2023public}, a free community-based platform. The topics of the visualizations are reflective of that demographic of users and include themes such as natural calamities, health, world events, financial news, entertainment, and sports, for example.

\begin{figure}
    \centering
    \includegraphics[width=\linewidth]{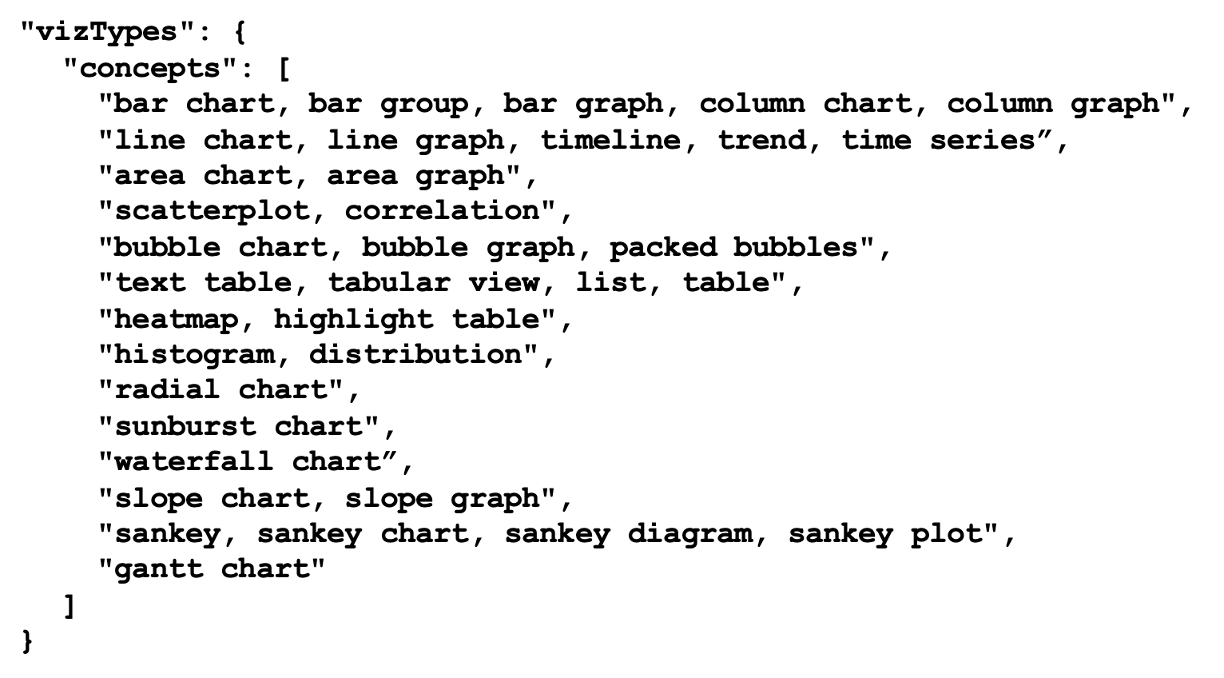}
    \caption{A JSON list of visualization types and their concepts that are stored as metadata to support design search.}
    \label{fig:viztypes}
\end{figure}

Given the XML visual specification of the Tableau workbooks, the system traverses the DOM structure and indexes any text metadata that can be extracted from the visualizations, similar to techniques described in~\cite{d3deconstruction}. Extracted metadata includes the visualization title, caption, tags, description, author name, and profile, the visualization marks encoded in the visualization, and the visualization type. To support design search for recognizing visualization types mentioned in the search query (\textbf{DC2}), we include a general list of visualization types and their linguistic variants in the semantic search framework, as shown in Figure~\ref{fig:viztypes}.

While we focused on CSV data sources and Tableau visualizations, the architecture for \olio~is extensible to include any new or additional data repositories, including D3 and Vega-lite charts, and knowledgebase articles, for example.

\noindent We now describe the rest of \olio's system components in detail.

\subsection{Query Classifier}
 \olio~takes as input an NL search query that is passed to the \textit{query classifier}. The classifier supports federated query search~\cite{Shokouhi2011FederatedS}, which is the process of distributing a query to multiple search repositories and combining results into a single, consolidated search result. Thus, for users, it appears as if they were interacting with a single search instance (\textbf{DC1}). In this context, a user can search \olio~over heterogeneous data repositories (i.e., both data sources and visualizations) without having to change or modify how they structure the query input. The query classifier passes the search tokens to a \emph{parser} and the \emph{data source search index} (which is part of the semantic search framework) and determines if ~\olio~needs to generate a Q\&A search to dynamically generate visualization responses, or simply general search that supports both exploratory and design searches. Algorithm 1 describes the query classification process. At a high level, the query classifier passes the query tokens to the parser (line 7) to determine if the query contains any analytic intents such as aggregation, correlation, temporal, or geospatial expression (refer to Section \ref{sec:parser} for more details). The query classifier also passes the query tokens to the semantic search framework (refer to Section \ref{sec:semanticsearch} for more details) to determine if the query tokens match fields in any of the data sources (e.g., `prices' $\rightarrow$ \texttt{Price} in the housing data source) and the normalized match score is greater than a predetermined threshold (line 10). In practice, we found that $fieldMatch = 2$ and $normMatch = .3$ provided a reasonable threshold for relevant data source matches. If both conditions, i.e., the presence of an analytical intent and the match score meets the threshold criteria, then Q\&A search is first invoked to dynamically generate visualization responses to the given query (line 13); else, general search is invoked to return pre-authored content from the data repository (line 16).

\begin{algorithm}
\caption{Classifies the search behavior based on whether the query contains an analytical intent and there is a match on one or more of the curated data sources in \olio.}
 \flushleft
\begin{algorithmic}[1]
\Function{QueryClassifier}{$query$}
    \LeftComment*{Boolean to check if there is an analytical intent in query}
    \State $hasAnalyticalIntent \leftarrow False$
    \LeftComment*{Boolean to check if there is a data source match}
    \State $hasDSMatch \leftarrow False$
    \LeftComment*{Contains the match scores for $query$ and each data source, $ds$}
    \State $dsScores \leftarrow getDSScores(query, ds)$
    \LeftComment*{Contains the normalized match scores for $query$ and each data source, $ds$}
    \State $normScores \leftarrow norm(dsScores)$ 
    \LeftComment*{Predetermined thresholds set for field match in $ds$ and $normScores$}
    \State $fieldMatch$, $normMatch$
     \LeftComment{Check if the parsed query contains an analytical intent}    
    \If{$(parseForAnalyticalIntent(query)$}
        \State $hasAnalyticalIntent \leftarrow True$  
    \EndIf
    \LeftComment{Check if the query tokens match fields in $ds$ and normalized match score to $ds$ is greater than a pre-determined thresold}     
    \If{$(dsScores['fields'] > fieldMatch)$ \textbf{and} $(normScores > normMatch))$}
        \State $hasDSMatch \leftarrow True$  
    \EndIf
    \LeftComment{If $query$ has an analytical intent and contains tokens matching a $ds$, invoke Q\&A search before general search, else just invoke general search.}
    \If{($hasAnalyticalIntent$ \textbf{and} $hasDSMatch$)}
        \State $invokeQ\&ASearch(query, ds)$
    \EndIf
        \State $invokeGeneralSearch(query)$
\EndFunction
\end{algorithmic}
\end{algorithm}

\begin{figure*}[ht]
    \centering
    \includegraphics[width=\linewidth]{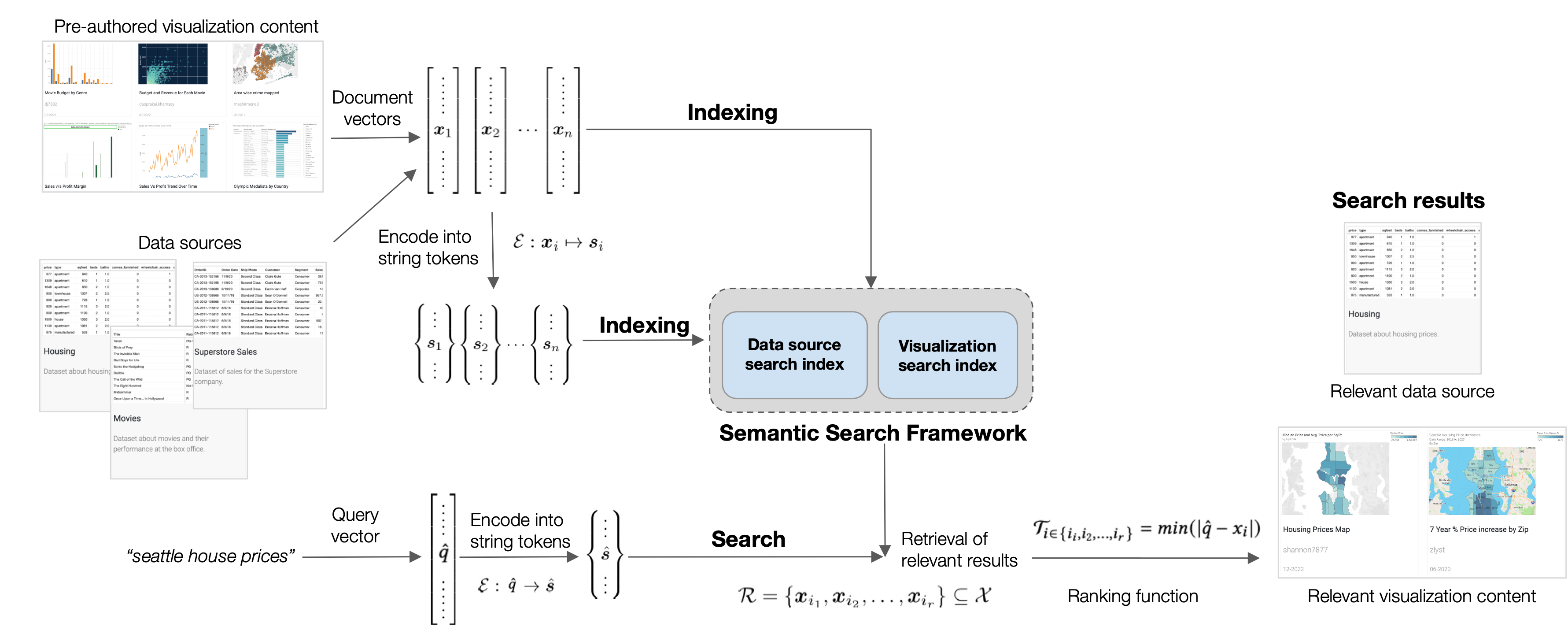}
    \caption{Pipeline of the semantic search framework. The document vectors, $\mathcal{X}$ from the pre-authored visualization content and data sources, along with their corresponding encoded string tokens, $\mathcal{S}$, are indexed in the semantic search framework as two data repository indices. At search time, the query vector, $\hat{q}$ from the input search query, \textit{``seattle house prices,''} is encoded into string tokens, and a set of relevant results are returned for both the visualization content and data sources. Using the ranking function, $\mathcal{T}_{i \in \{i_i, i_2,...,i_r\}} = min(|\hat{q} - x_i|)$, the result set is finally ranked to return the top-scoring results.}
    \label{fig:semanticsearchframework}
\end{figure*}

\subsection{Parser}
\label{sec:parser}
The parser removes stopwords (e.g., `a', `the') and conjunctions / disjunctions (e.g., `and,' `or') from the search query and extracts a list of n-gram tokens (e.g., ``\textit{Seattle house prices}'' $\rightarrow$ [Seattle], [house], [prices], [house prices], [Seattle house prices], etc.). The parser employs a Cocke-Kasami-Younger (CKY) parsing algorithm ~\cite{cocke1969programming,kasami1966efficient,younger1967recognition} and generates a dependency tree to understand relationships between words in the query. 

The input to the underlying CKY parser is a context-free grammar with production rules augmented with both syntactic and semantic predicates to detect the following analytical intents in the search query:

\begin{tight_itemize}
\item \textbf{Grouping}. Partition the data into categories. E.g., `by' a data attribute.
\item \textbf{Aggregation}. Group values of multiple rows of data together to form a single value
based on a mathematical operation. E.g., `average,' `median,' `count,' `distinct count.'
\item \textbf{Correlation}. Statistical measure of the strength of the relationship between two data attributes (measures). E.g., `correlate,' `relate.'
\item \textbf{Filters and limits}. Finite sets of operators that return a subset of the data attribute's domain. E.g., `filter to,' ‘at least,’ ‘between,’ ‘at most.’ Limits are also a finite set of operators akin to filters that return a subset of the attribute’s domain, restricting up to n rows. E.g., `top,' `bottom.'
\item \textbf{Temporal}. Time and date expressions containing temporal tokens and phrases. E.g., `over time,' `year,` `in 2020', `when.'
\item \textbf{Geospatial}. Geospatial expressions referring to location and place. E.g., `in Canada,'  `by location,' `where.'
\end{tight_itemize}

To help with detecting data attributes and values along with the intents, the parser has access to the set of curated data sources and their metadata. The parser then compares the n-grams to available data attributes looking for both syntactic (e.g., misspellings) and semantic similarities (e.g., synonyms) using the Levenshtein distance~\cite{yujian2007normalized} and the Wu-Palmer similarity score~\cite{wu1994verbs}, respectively (\textbf{DC2}). If the parser detects one or more of the aforementioned analytical intents, it returns the intent(s) along with its corresponding data attributes and values to the query classifier.

\subsection{Semantic Search Framework}
\label{sec:semanticsearch}
The semantic search framework primarily comprises two phases: indexing and searching content and metadata in the data repositories. This two-phase process applies to content in the data repositories, i.e., both the curated data sources and visualization content. Figure~\ref{fig:semanticsearchframework} illustrates the pipeline of the semantic search framework.

\subsubsection{Indexing}
The indexing phase creates indices for each of the data repositories (data sources and visualization content) along with their metadata to support federated search in \olio~(\textbf{DC1}).

Given a data source and visualization content with associated metadata (i.e., attributes, data values, chart type, author name), each file is represented as a document vector, $x_i$, where:
\useshortskip
\begin{equation}
    \mathcal{X} = \{x_1, x_2, ..., x_n\} 
\end{equation}

We also store n-gram string tokens from these document vectors to support partial and exact matches in the system (\textbf{DC2}):
\useshortskip
\begin{equation}
\mathcal{S} = \{s_1, s_2, ...s_n\}
\end{equation}

where $s_i = \varepsilon(x_i)$ for some encoder, $\varepsilon$ that converts the document vectors into a collection of string tokens of cardinality $n$. The original vectors $\mathcal{X}$ and encoded tokens $\mathcal{S}$ are stored in the semantic search engine index by specifying the \emph{mapping} of the content, i.e., defining the type and format of the fields in the index. \olio~stores the text as keywords in the index, supporting exact-value search, fuzzy matching to handle typos and spelling variations, and n-n-grams for phrasal matching. A scoring algorithm, tokenizers, and filters are specified as part of the search index \emph{settings} to determine how the matched documents are scored with respect to the input query and the handling of tokens, such as the adding of synonyms from a thesaurus, removal of stopwords (e.g., `a,' 'the,' for') and duplicate tokens, and converting tokens to lowercase. The complete configuration specification is provided in supplementary material.


\subsubsection{Search}
\label{sec:search}
Conceptually, the search phase has two steps: retrieval and ranking. Given an input query, $q$, that is represented as a query vector, $\hat{q}$ with query tokens $q_1, q_2, ..., q_j$; we encode the vector into string tokens, $\hat{s} = \varepsilon(\hat{q})$ using the same encoder, $\varepsilon$ from the indexing phase. The search process retrieves the most relevant $r$ document vectors, $\mathcal{R} = \{x_{1}, x_{2}, ... x_{r}\}$ as candidates based on the amount of overlap between the query string token set $\hat{s}$ and the document string tokens in $\{s_1, s_2, ..., s_n\}$. More specifically, the scoring function $r_{max}$ maximizes search relevance by computing:
\useshortskip
\begin{equation}
    \{x_1, x_2, ..., x_r\} = {r_{max}}_{i \in \{1, 2, ..., n\}} |\hat{s} \cap s_i |
\end{equation}

\noindent\olio~then ranks the vectors in the candidate search result set, $\mathcal{R}$ based on $BM25$ scoring~\cite{manning2008introduction} with respect to the query vector, $\hat{q}$. BM25 is essentially a bag-of-words retrieval scoring function that ranks documents based on the query terms appearing in each document, regardless of their proximity within the document. It is a preferred metric for computing similarities between vectors as the method corrects for variations in vector magnitudes resulting from uneven-length documents~\cite{manning2008introduction}. Given  $\hat{q}$, the BM25 score of a document vector, $x_i$ is:

\useshortskip
\begin{equation}
     {BM25}(\hat{q}, x_i) = \Sigma^{n}_{i=1}IDF(q_j).\frac{f(q_j, x_i).(k_1 + 1)}{f(q_j, x_i) + k_1. (1 - b + b . \frac{|x_i|}{avgdl})}
\end{equation}

where $f(q_j, x_i)$ is the number of times that  $q_{j}$ occurs in the document vector, $x_i$ and $avgdl$ is the average document vector length in the search index. $k_{1}$ and $b$ are constants to further optimize the scoring function. In practice, we have found that $k_1 \in [1.2,2.0]$ and $b = 0.75$ tend to provide reasonable ranking behavior. The Inverse Document Frequency, $IDF$, measures how often a term occurs in all of the documents and ranks unique terms in documents higher. It is computed as:

\useshortskip
\begin{equation}
IDF = ln(1 + \frac{(docCnt - f(q_j) + 0.5)}{f(q_j) + 0.5}
\end{equation}

where $docCnt$ is the total number of documents that have a value for the given query token, $q_j$ and $f(q_j)$ is the number of documents that contains the $i^{th}$ query term.

The $BM25$ scoring function sorts the vectors in descending order of normalized $BM25$ scores, $b \in [0,1]$, i.e., the higher the score, the higher the rank, creating the final ranked search result set,  $\mathcal{T}$, ranked based on the minimum difference between the query and each of the document vectors:

\useshortskip
\begin{equation}
    \mathcal{T}_{i \in \{i_i, i_2,...,i_r\}} = min(|\hat{q} - x_i|)
\end{equation}


The search request is then passed to the Elasticsearch server to compute Equations 3 and 4 and the system returns a ranked result set of either data sources (used for Q\&A) or visualization content used for both exploratory and design search scenarios.

\subsection{Q\&A Module}

\begin{figure}
    \centering
    \includegraphics[width=.8\linewidth]{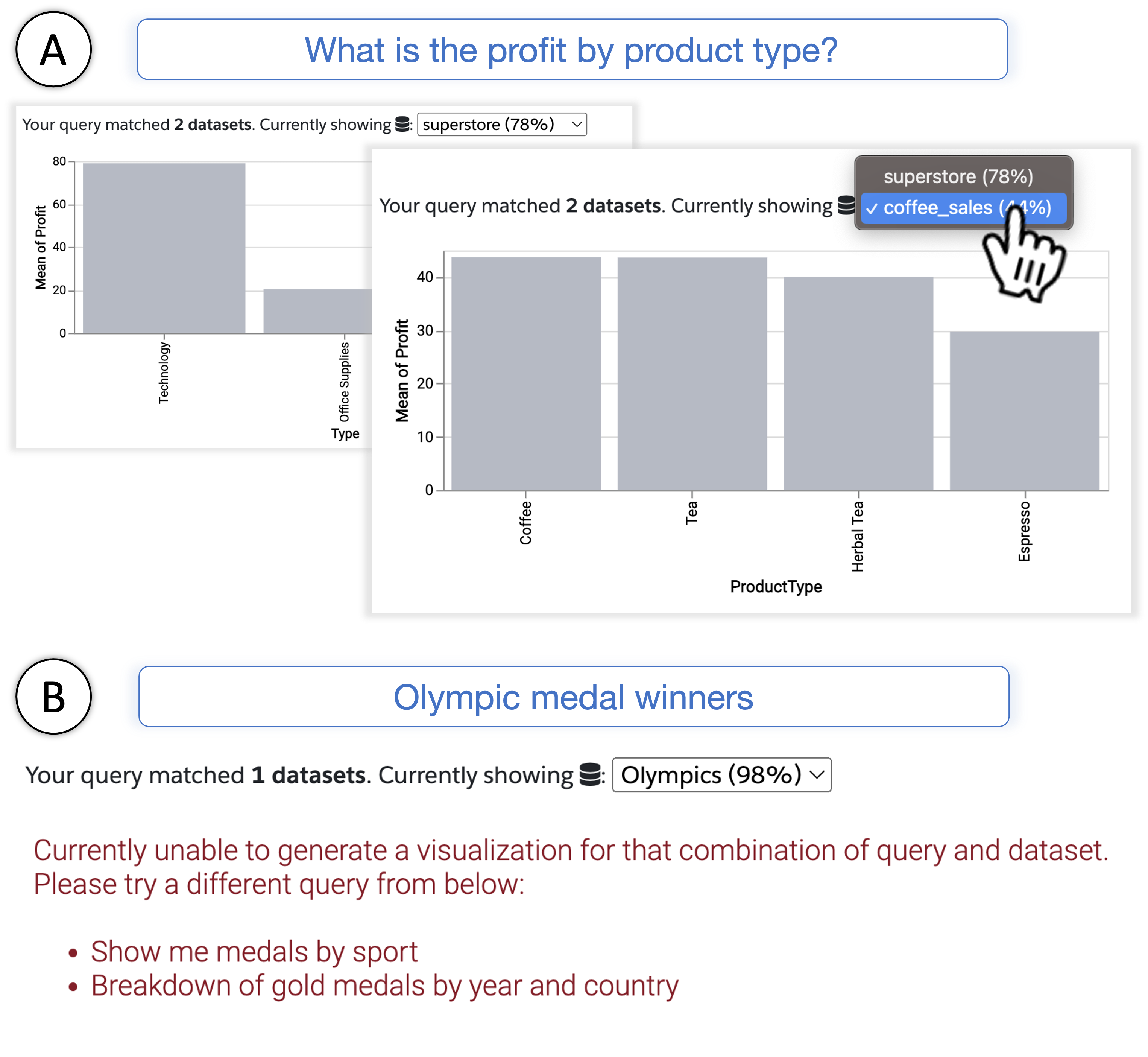}
    \caption{The Q\&A portion of the interface in \olio~provides interaction and scaffolding support  (\textbf{DC3}). (A) In the case that multiple data sources are identified as top matches to a given search query, a drop-down list shows the ranked list of data sources along with their corresponding percentage match scores. (B) In the case that no valid visualization can be generated for a given search query even though there is a match to a data source, \olio~displays a list of query suggestions that the user could choose from to generate a visualization response.}
    \label{fig:qa-special-cases}
\end{figure}

The \emph{Q\&A module} interprets the analytical intent expressed in the input search queries and dynamically generates visualization responses based on the list of top-matched data source(s) returned from the semantic search framework, as described in Section~\ref{sec:semanticsearch}. The module accepts tabular CSV datasets for the top-matched data source(s) as input, and all the visualizations in the tool are created using Vega-Lite~\cite{satyanarayan2016vega} and D3~\cite{2011-d3}. 

The interface and functionality for Q\&A search in \olio~is similar to that of NLIs for visual analysis~\cite{datatone,eviza,orko} with a few extensions that are inherent to the Q\&A behavior in the context of semantic search. For instance, the interface displays text showing a match (if any), to one or more data sources, along with a drop-down menu of the matched data sources  (\textbf{DC3}). A visualization is rendered based on attributes, values, and the analytical intent in the query, along with a text summary describing the visualization (refer to Figure~\ref{fig:teaser}A). A user can peruse the drop-down list of other data source alternatives, along with their corresponding percentage match scores (as computed in Section~\ref{sec:search}), and choose to switch to another data source in the drop-down list as shown in Figure~\ref{fig:qa-special-cases}A. In cases where there is a match to a data source for the query, but the tokens in the query do not resolve to valid attributes and values within the data source, \olio~displays suggested queries for the data source (\textbf{DC3}), shown in Figure~\ref{fig:qa-special-cases}B. These query suggestions are generated using a template-based approach presented by Srinivasan and Setlur~\cite{srinivasan2021snowy} that is based on a combination of attributes from the data source and data interestingness metrics.

The visualization generation process for Q\&A search supports three encoding channels (\texttt{x}, \texttt{y}, \texttt{color}) and four mark types (\texttt{bar}, \texttt{line}, \texttt{point}, and \texttt{geoshape}). These marks and encodings support the dynamic generation of bar charts, line charts, scatterplots, and maps that cover the range of analytic intents described in Section~\ref{sec:parser}. \olio~selects the default visualization using a simplified version of the Show Me system~\cite{mackinlay2007show}, employing similar rules to determine mark types based on the mappings between the visual encodings and attribute data types (e.g., showing a scatterplot if two quantitative attributes are mapped to the \texttt{xy}-channels and showing a line chart if a temporal attribute is visualized on the \texttt{x}-axis with a quantitative attribute on the \texttt{y}-axis).

Finally, \olio~displays a dynamic text summary describing the generated visualization (\textbf{DC3}). While template-based approaches~\cite{kim:2021,fasciano-lapalme-1996-postgraphe,mittal:1995} are viable options for the summary generation process, we chose to employ a large language model (LLM)-based approach~\cite{chatgpt} to explore its capabilities and better understand its limitations.
We initially attempted to pass the chart data as-is to ChatGPT to generate a description. However, we found the model was oftentimes generating wrong statistics or even hallucinating depending on the data domain context. To overcome these challenges but still provide an eloquent description, we instead opted for a combined approach of using both basic statistical computations and an LLM.

Specifically, the input to ChatGPT is a prompt containing a statistical description that is extracted from the generated visualization using a set of heuristics defined in prior data insight recommendation tools~\cite{demiralp2017foresight,cui2019datasite,srinivasan2018augmenting}.
For instance, for bar charts, we identify the min/max and average values; for scatterplots, we compute the Pearson's correlation coefficient~\cite{freedman2007statistics}, and so on. Consider the search query, \textit{``sales by region,''} which results in a bar chart displaying \texttt{Sales} across four \texttt{Region}s. An example of the statistical description, $keyStats$ from this bar chart is:
\begin{verbatim}
Region: Central has a minimum value of $220 for Sales
Region: South has the maximum value of $240 for Sales
Average Sales across Region is: $230
\end{verbatim}

\noindent The corresponding prompt to ChatGPT then becomes \textit{Rephrase the following input more eloquently: \escape{n}`\$\{keyStats\}\escape{n}'}, which ultimately generates the text summary: \textit{``The Sales in Central Region had the lowest value of \$220,  while South Region had the highest value of \$240. The average Sales across all Regions was \$230.}

\subsection{General Search Module}
\begin{figure}[t!]
    \centering
    \includegraphics[width=\linewidth]{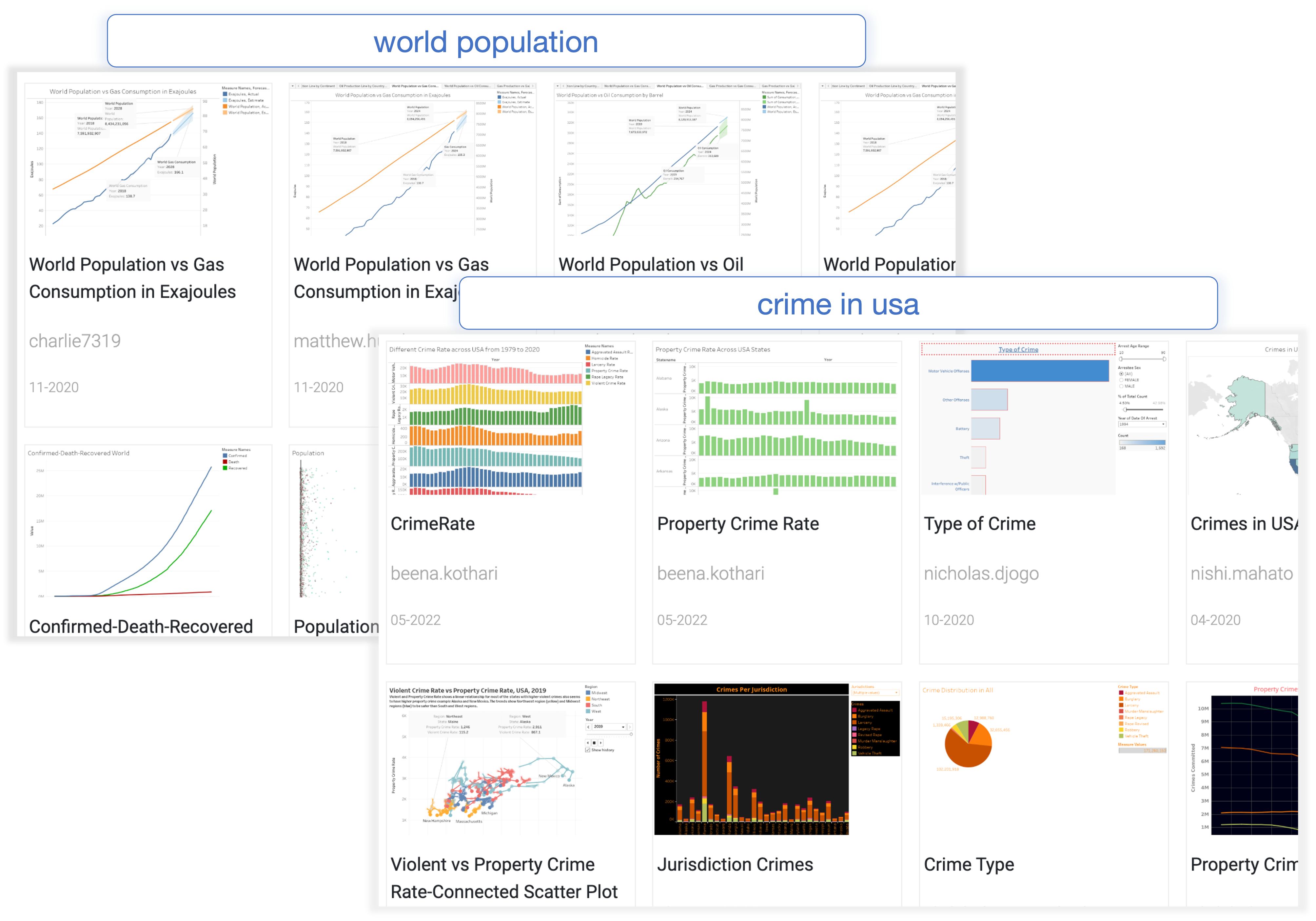}
    \caption{Exploratory search examples. \olio~displays thumbnail results of pre-authored visualizations based on keywords found in the input search queries, \emph{``world population''} and \emph{``crime in usa.''}}
    \label{fig:exploratory-search-examples}
\end{figure}

\begin{figure}[t!]
    \centering
    \includegraphics[width=\linewidth]{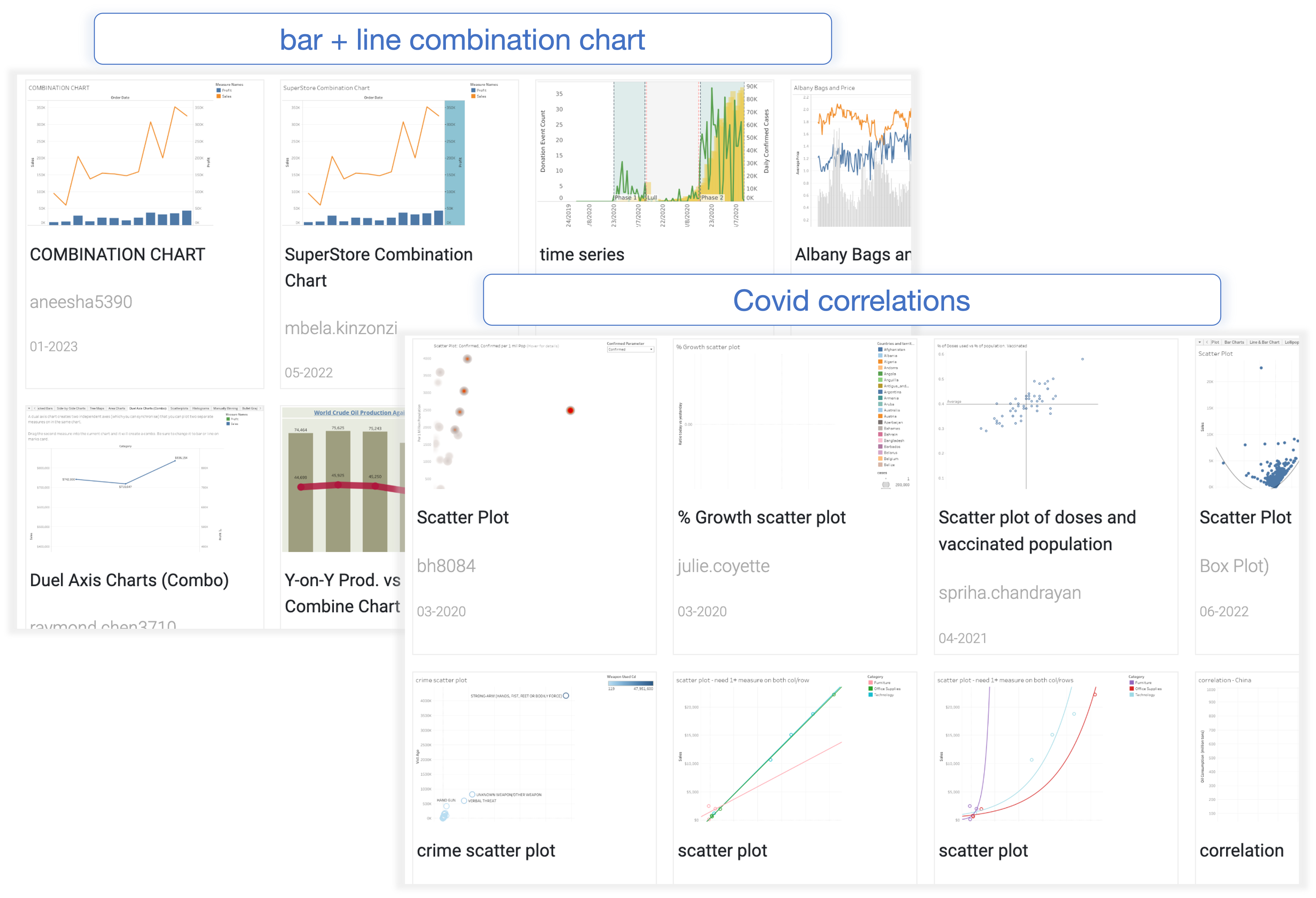}
    \caption{Design search examples. \olio~displays thumbnail results of pre-authored visualizations for combinations of chart types (e.g., bar and line charts) as well as for analytical concepts that allude to a specific visualization type (e.g., `correlation' for scatterplot).}
    \label{fig:design-search-examples}
\end{figure}

The \emph{general search module} displays thumbnails of pre-authored visualization content along with information such as title and date. The thumbnail images are hyperlinked to the corresponding Tableau Public workbook URLs if users choose to download or analyze the visualization in more detail. The module enables two types of searches: exploratory and design (\textbf{DC1}). Exploratory search returns visualization results based on keyword matches (\textbf{DC2}) in the input search query (e.g., \textit{``world population''} in Figure~\ref{fig:exploratory-search-examples}). Design search is a special form of exploratory search that returns visualization results specifically for keywords containing tokens referring to visualization types, their synonyms, and related concepts (e.g., \textit{``covid correlations''}) (\textbf{DC2}). Figure~\ref{fig:design-search-examples} shows examples of design search results in \olio.

\section{Preliminary User Study}


Using \olio{} as a design probe, we conducted a preliminary user study to qualitatively assess the overarching idea of combining dynamically generated visualizations with pre-authored charts when searching data repositories. \change{Note that while a comparison of \olio{} with other systems would be helpful in identifying their relative strengths and weaknesses, the current state-of-the-art semantic search engines~\cite{Bing,google,ding:2005} focus on web documents rather than data repositories. In addition, existing visualization search systems~\cite{hoque2019searching,sechler2017sightline} focus on a subset of search functionality supported in \olio{}. Removing individual components for an ablation study would be challenging due to \olio{}'s unified hybrid search behavior. However, \olio{} does implement industry-standard performant recommendations like BM25 scoring and ElasticSearch indexing.}

\subsection{Participants and Setup}

We recruited $11$ participants (P1-P11, \change{6 males and 5 females}) through a mailing list at a data analytics software company.
Based on self-reporting by the participants, five participants frequently searched for data or visualization content on data repositories, four participants had some experience with searching data repositories but did so infrequently, and two participants had minimal experience with search in the context of data and visualizations.

All sessions were conducted remotely via the Cisco WebEx video conferencing software~\cite{webex}.
The prototype was hosted on a local server running on the experimenter’s laptop\footnote{2.4 GHz MacBook Pro running macOS Ventura 13.2.1 set to a resolution of 3072 $\times$ 1920.}.
Participants were granted control over the experimenter’s screen during the session, and all studies followed a think-aloud protocol.
The audio, video, and on-screen actions were recorded for all sessions with permission from the participants.

\subsection{Procedure}

Sessions lasted between 39-60 minutes (mean: 46 min.) and were organized as follows:

\pheading{Introduction} [$\sim$10min]: After providing an overview of the study goal, the experimenter asked participants about their job roles and prior experience with search, particularly in the context of data and visualization. The participants were provided a brief introduction to \olio{'s} interface, highlighting the four key components listed in Figure~\ref{fig:interface}. \change{Consistent with Jeopardy-style evaluations of prior NLIs for visualization~\cite{datatone}, to} avoid biasing participants, we did not provide any explicit training or queries and instead allowed participants to implicitly discover the system through the study tasks.

\pheading{Task Phase} [$\sim$25min]: Participants were asked to perform four tasks: one task corresponding to each search goal listed in Section~\ref{sec:formative} and a fourth open-ended task where participants were allowed to freely explore the available data sources and pre-authored visualizations.

For the \textit{Q\&A} task, participants were asked to use one or more of the available data sources for a Jeopardy-style fact~\cite{datatone} about college admissions and a directed analysis question of ``listing 1-3 insights on differences between movie genres.''

For \textit{exploratory} search, participants were asked to use \olio{} to explore the topics of elections and colleges in the US. Participants were encouraged to use any search terms and phrase queries however they saw fit.

To assess \olio{'s} support for \textit{design} search, participants were given two images and were asked to search for similar examples using the tool. 
The images included a treemap showing stock data and a choropleth map of US states with an overlaid pie chart showing product sales data.

\pheading{Debrief} [$\sim$10min]: Sessions concluded with a semi-structured interview discussing the overall experience and utility of the underlying idea, support for different search goals, and areas for improvement.

\subsection{Results}

Overall, participants noted that the semantic search paradigm was useful and could help accomplish their search goals in the context of data repositories. Below, we detail participant feedback and usage behavior with respect to the three search goals listed in Section~\ref{sec:formative}.

\pheading{Q\&A.}
All participants successfully completed the two Q\&A tasks and generally appreciated the system's ability to interpret different phrasing variations (e.g., \textit{``tuition across us regions,''} \textit{``compare movie genres,''} \textit{``What were covid cases across countries?'\''}).
P9, for instance, said, ``\textit{I think the system did better than what I would expect in terms answering questions even though my questions were not good enough to begin with.}''
Participants also used the system's ability to dynamically generate visualizations for in-place data exploration.
For example, P3 issued a query, \textit{``What are movie budgets by genre?''} that resulted in a bar chart showing average \texttt{Budget} by \texttt{Genre}.
Then, using the metadata tooltip (Figure~\ref{fig:start-screen}), he inspected other fields to notice the \texttt{Gross} field and issued a query to visualize both \texttt{Budget} and \texttt{Gross}.
P6, P7, and P9 also exhibited a similar behavior on multiple instances suggesting that the dynamic content promoted a state of analytic flow.
Participants also appreciated the ability to view and choose from matched data sources (Figure~\ref{fig:qa-special-cases}A).
Specifically, participants commented that \olio{} provided the freedom to ``\textit{get more with less}'' by supporting keyword-based or open-ended queries to retrieve multiple data sources instead of focusing on well-phrased queries that were optimized to match a single data source.
Proposing an improvement to the current interface, however, P10 suggested that instead of rendering a visualization by default, when there are multiple data source matches, the system could allow first choosing a data source and then rendering the chart to save computation resources at scale.

\pheading{Exploratory search.}
Participants commented that \olio{} returned appropriate sets of pre-authored charts during open-ended exploratory searches (e.g., `elections,' `olympics winners,' `covid trends').
However, we noticed that the quality of search results deteriorated when queries went beyond keywords and included additional information such as location (e.g., `election results in Maryland'), subjective concepts (e.g., `safest cities in the us'), or metadata properties like `popularity' that were not included in our chart corpus (e.g., `popular NBA charts').
Although there were mixed reactions to the quality of search results for exploratory scenarios, all participants appreciated the form and function of the dynamic filtering widgets (Figure~\ref{fig:interface}D), commenting they ``\textit{loved it}'' (P8, P11) and asked ``\textit{why these [dynamic filtering widgets] don't exist in all systems today?}'' (P10).
Participants predominantly used filters to facet the search results (e.g., choosing chart types to focus on a subset of results or specifying a time range to focus on recent results).
On two occasions, participants (P6, P11) also leveraged the filters to chronologically compare search results.
When exploring the topic of `us elections,' for instance, P11 used the date range slider widget to focus on charts created during the 2020 elections to those created using the 2016 elections.

\pheading{Design search.}
All participants successfully completed the two design search tasks except for P1 and P5, who found only one of the two required charts.
Overall, participants were very positive about the system's support for searching for visualizations by design features, with P8 stating, \textit{``I would love to have this in Tableau Public today.''}
Similarly, P7 also noted, \textit{``if all we do from this system is enable this search by design [in other visualization repositories], I'd argue that it can solve a lot of challenges for chart authors and especially for someone new to visualization tools.''}
In terms of user behavior, as we expected, some participants (6 out of 11) did not recollect `treemap' as a chart type and instead used the data topic, the closest chart type they could think of, or the mark type (e.g., `stock heatmap,' `finance group blocks,' `square chart').
However, since \olio{} inspects the content of the chart (e.g., titles) as well as design features (e.g., chart type, mark type), the system was able to return relevant charts as some of its top results.
In combination with the chart type filter widget, this feature enabled participants to reliably find example charts even when they could not precisely describe them.
Participants also successfully used a variety of phrasings to find the combined map and pie chart (e.g., \textit{``examples of piemaps,''} \textit{``pie+map,''} \textit{``show me charts with sales on a map with overlaid pie charts''}).
\section{Discussion}
\label{sec:discussion}

Besides user feedback on \olio{'s} support for different search scenarios, the study also helped identify high-level themes and user behaviors pertaining to the semantic search paradigm.

\pheading{Hybrid results facilitate a fluid and analytical search experience.}
When talking about the utility of the presented idea, participants particularly appreciated the \textit{complementary nature} of the dynamically generated content and the pre-authored visualizations.
Noting the benefits of each component, P9, for instance, said, ``\textit{if there's a question that can be answered using a data source then dynamically generated content like this is going to save a lot of time... But when I'm looking for inspiration, of course, that's not the best way, and what I would look for is work by actual people so definitely I see applications for both. None of them are mutually exclusive, and I was able to utilize both of them.}''
P2 viewed pre-authored content as a fail-safe for cases when there is no dynamic content stating, ``\textit{even if you don't have a dataset that's directly relevant to your query, if there are visualizations, then they come up immediately, which I really appreciate.}''
We also observed that the combination of the two content types encouraged participants to introspect on the data and findings more closely.
For example, P11 issued a query, \textit{``compare movies by genre''} that generated a bar chart from one of the available data sources, depicting that the \texttt{Action} genre has the highest number of movies.
However, she found a similar chart in the pre-authored set that showed a different result and correspondingly started inquiring about the data source, what dates it covered, if certain movies were excluded, etc.

\pheading{The link (or the lack thereof) between the dynamic chart and the pre-authored content should be more apparent.}
Although participants understood the differences between the two types of results, some participants were initially confused that the dynamic and pre-authored content did not stem from the same data source.
P7 alluded to this initial confusion about the visual layout of the page, stating that ``\textit{the page kind of creates a hierarchy that is difficult to break. I thought that there was the data I'm looking at at the top was getting visualized in different ways at the bottom, and that was that.}'' P2 suggested adding a button above the filters in the interface (Figure~\ref{fig:interface}D) to toggle the pre-authored results to only those that are created using the same data source as the dynamic result.
This feedback suggests that for the semantic search experience to be effective, systems like \olio{} should explore interface designs that clearly depict the relationship between content types, providing users the option to update the content ad-hoc.

\pheading{The inclusion of dynamically generated content changes user expectations.}
We noticed an intriguing change in the querying pattern for some participants (P3, P5, P7, P11) as they became familiar with the tool and experienced dynamic content as part of the results.
Specifically, once the system generated charts for a few queries, they switched from treating \olio{} as a search tool using keyword-style queries as input to more of an NLI, issuing imperative system commands like \textit{``Show me a chart of tuition cost by region''} and \textit{``Display examples of treemaps showing stock market data.''} 
While \olio{'s} query parsing logic was able to accommodate most phrasing variations, there were cases where the system no longer met the participants' expectations.
For instance, P3 issued a query, \textit{``show examples of charts displaying sales by state''} and \olio{} returned a map and bar chart for the Superstore data sources as part of its dynamic content along with other pre-authored charts matching the search query.
However, P3 was confused by this result as he expected the system to understand the phrase \textit{`show examples'} and ignore the data source search and dynamic chart rendering altogether.
When asked about the change in their querying patterns during the session, multiple participants (P3, P11) commented that it was a combination of \olio{} initially exhibiting an understanding of well-formed natural language utterances and their recent exposure to a slew of conversational interaction experiences through language models like ChatGPT.
Such mismatches in the system's functionality (supporting search) and the user's expectation (conversational interaction with an agent) could lead to errors in a larger scale setting, however, and should be clarified through a combination of interface techniques and system guidance.

\pheading{Textual descriptions should provide structure and contextual information.}
Participants' reactions to the system-generated descriptions were lukewarm at best, with only four participants (P4, P7, P9, and P10) commenting on them during the study.
During their comments, participants noted that the text was helpful in that it re-iterated the key facts from the chart, making it easy to interpret the chart, particularly when it was very dense with overlapping marks (e.g., a multi-series line chart or a scatterplot).
However, participants felt that ``\textit{text structure is too verbose}'' (P7) and ``\textit{lacks contextual information about what it means for a value to be high or low},'' (P11) minimizing its overall utility.
Such comments suggest that future systems investigating text generation in the context of data repository search should not only focus on the mapping between the generated text and chart, but also on the structure and degree of external information in the text itself.
\section{Limitations and Future Work}

Semantic search interfaces for data repositories hold promise for helping a user navigate and explore the growing amount of visualizations and analytical assets available. While \olio~received positive feedback as a research probe, research exploring semantic search for data repositories is still in its infancy. We identify various themes that highlight the challenges and opportunities for supporting semantic search that are unique to data repositories.

\pheading{Search precision depends on the availability and curation quality of data sources.} \change{Similar to other semantic search experiences~\cite{kaufmann:2006,klein-manning-2003-accurate,KOLOMIYETS20115412}, Q\&A search utilizes a small set of curated datasets to address analytical intents with focused responses.} However, an important aspect of search precision, especially for dynamically generated responses, is having access to high-quality, curated data sources with well-understood semantics. However, there is often a disconnect between environments where users publish content and downstream applications like search that consume the content. Participants echoed this challenge with P10 stating, ``\textit{this is a great interface and experience but will have to overcome the data garbage problem at scale}.''
Authors tend to perform some amount of curation during the publishing process but often are not provided sufficient tools to annotate, tag, or enrich their content. The process of curation is often tedious and time-consuming. More research should explore techniques (both semi-automated and automated)~\cite{potterswheel,wrangler,datacleaning:survey} to reduce the friction while curating content in data repositories; this includes the de-duplication of similar or near-similar content and the suggestion of topics and tags to help with content discoverability and faceting. \change{Future work should also explore techniques to help with data curation, such as employing LLMs for metadata enrichment, incorporating entity recognition, synonyms, and relational extraction to help automate curation for Q\&A support.}

\pheading{Incorporating additional analytical assets and metadata.} \olio~\\currently searches over pre-authored singleton visualizations. Future extensions should consider expanding the repertoire of analytical assets to include dashboards, data tables, and computational notebooks~\cite{Observable,jupyter}. These forms of content have interesting implications for interpreting analytical intent, Q\&A, and design search beyond data source and visualization repositories. Further, combining data repositories with document repositories could provide additional searchable metadata to improve search precision and for generating contextually relevant summaries alongside the results.

\pheading{Need for scaffolding to orient the user.}
Semantic search interfaces support new techniques for information seeking but with the added complexity of determining the type of queries and understanding the search results. Guidance and scaffolding may need to be provided as users search across multiple data repositories of content. While \olio~displays metadata for the available data sources along with query suggestions to guide a user toward a successful search, additional scaffolding could improve sensemaking and exploration. Recent work has explored data-driven autocompletion for helping users formulate targeted Q\&A-type queries~\cite{sneakpique} and integrate contextual query suggestions within a person’s sensemaking environment~\cite{interweave}. An interesting research direction would be to explore data scaffolds across different types of search, each unique in its own way, in the context of a semantic search system.

\pheading{Explore new search paradigms and modalities.} \olio~indexes available textual content in the data repositories. However, akin to image search, content-based search~\cite{cbr:2000} that leverages \emph{visual} features could improve recall of sparse text content, particularly for design search. Reverse image search~\cite{visualsearchpinterest} addresses the challenge for a user to guess at keywords and terms to return a specific result that they may have in mind. Exploring reverse visualization search, wherein a user provides a sample visualization or sketch to discover content related to the sample visualization image, could support richer forms of expressing design search goals. In addition to new search paradigms, other modalities, and platforms should be explored. Mobile devices, for example, generate large amounts of sensor footprints (e.g., GPS, motion sensors) and user activity data that are often missing from their desktop counterparts~\cite{franti2005mobile}. These new sources of implicit and explicit user feedback are valuable for discovering actionable content which is both situationally and contextually relevant to the user. Further, voice and touch modalities could open new possibilities for query formulation and browsing content in the data repositories.

\pheading{Trust and provenance.} Trust is an important issue, and users would benefit from information that communicates the provenance of data sources used to generate the visualization responses, along with the ranking of pre-authored content. Exploring the inclusion of explanations for the search results could lead to increased transparency and understanding of the system behavior~\cite{ramos:2020}. There are additional challenges in an enterprise context; data and visualization content may be private to certain teams and organizations due to the sensitivity of the data (e.g., a human resources department or the current revenue forecast of a business). More work needs to explore ways to support built-in data privacy for indexing and searching of content within these organizational boundaries.

\pheading{Exploring the utility of LLMs for search.} Due to their ease of use and their fluent text-generative capabilities, LLMs are garnering attention for search and conversational interfaces~\cite{meyer:2022}. We explored the use of ChatGPT to generate a summary of the dynamically generated visualization response for Q\&A. The model does have limitations in the types of summaries it can generate (as described in Section~\ref{sec:discussion}) and challenges around higher-order numeracy reasoning~\cite{frieder2023mathematical}. Custom-trained GPT models could potentially bridge this gap in higher-order analytical reasoning if they can be trained on the data repositories employed in a semantic search system. In addition to summary generation, other utilities for these custom LLMs could explore automatic metadata generation from data repositories to enrich sparse searchable text content. Understanding the quality and accuracy of the generated text both for metadata ingestion and summary generation\change{, and comparing the resulting search experience to that of \olio{}}, are important research directions to pursue as future work.


\section{Conclusion}
 In this paper, we explore how we can support data sensemaking and exploration in a semantic search paradigm designed specifically for data repositories. We introduce \olio, a research probe that realizes semantic search behavior through three types of searches: Q\&A, exploratory, and design. The system implements a novel semantic search framework that leverages analytical intent derived from the user's query, along with searchable metadata and content to provide a hybrid set of dynamically generated visualization responses with pre-authored visualizations. A preliminary evaluation of \olio~indicates that users find the system helpful for supporting a range of both targeted and open-ended data exploration activities. As people continue to actively explore data and author visualizations, there will be an increasing amount of searchable analytical content made available in these data repositories. The ability to support more expressive ways to utilize the content for a wide range of search goals will become especially important. This work provides interesting opportunities for managing and interacting with data beyond search; data curation and enrichment, along with novel modalities for exploring more varieties of content can further scaffold analytical discovery and insights.


\bibliographystyle{ACM-Reference-Format}
\bibliography{main}


\begin{thebibliography}{112}


\ifx \showCODEN    \undefined \def \showCODEN     #1{\unskip}     \fi
\ifx \showDOI      \undefined \def \showDOI       #1{#1}\fi
\ifx \showISBNx    \undefined \def \showISBNx     #1{\unskip}     \fi
\ifx \showISBNxiii \undefined \def \showISBNxiii  #1{\unskip}     \fi
\ifx \showISSN     \undefined \def \showISSN      #1{\unskip}     \fi
\ifx \showLCCN     \undefined \def \showLCCN      #1{\unskip}     \fi
\ifx \shownote     \undefined \def \shownote      #1{#1}          \fi
\ifx \showarticletitle \undefined \def \showarticletitle #1{#1}   \fi
\ifx \showURL      \undefined \def \showURL       {\relax}        \fi
\providecommand\bibfield[2]{#2}
\providecommand\bibinfo[2]{#2}
\providecommand\natexlab[1]{#1}
\providecommand\showeprint[2][]{arXiv:#2}

\bibitem[rgr(2018)]%
        {rgriffin_2018}
 \bibinfo{year}{2018}\natexlab{}.
\newblock \bibinfo{title}{120 years of {O}lympic history: Athletes and
  results}.
\newblock
  \bibinfo{howpublished}{\url{https://www.kaggle.com/datasets/heesoo37/120-years-of-olympic-history-athletes-and-results}}.
\newblock
\newblock
\shownote{Accessed: 2023}.


\bibitem[Bin(2023)]%
        {Bing}
 \bibinfo{year}{2023}\natexlab{}.
\newblock \bibinfo{title}{Bing {S}earch}.
\newblock \bibinfo{howpublished}{\url{https://www.bing.com/}}.
\newblock


\bibitem[web(2023)]%
        {webex}
 \bibinfo{year}{2023}\natexlab{}.
\newblock \bibinfo{title}{{C}isco {W}ebex\textsuperscript{TM}}.
\newblock \bibinfo{howpublished}{\url{https://www.webex.com}}.
\newblock


\bibitem[cov(2023)]%
        {covid-ca}
 \bibinfo{year}{2023}\natexlab{}.
\newblock \bibinfo{title}{Covid-19 {D}ataset}.
\newblock
\newblock
\newblock
\shownote{CC-BY Dataset: \url{https://covid19.ca.gov}}.


\bibitem[ela(2023)]%
        {elasticsearch}
 \bibinfo{year}{2023}\natexlab{}.
\newblock \bibinfo{title}{{E}lasticsearch}.
\newblock \bibinfo{howpublished}{\url{https://www.elastic.co/elasticsearch/}}.
\newblock


\bibitem[goo(2023)]%
        {google}
 \bibinfo{year}{2023}\natexlab{}.
\newblock \bibinfo{title}{Google {S}earch}.
\newblock \bibinfo{howpublished}{\url{https://www.google.com/}}.
\newblock


\bibitem[ibm(2023)]%
        {ibmwatson}
 \bibinfo{year}{2023}\natexlab{}.
\newblock \bibinfo{title}{{IBM} {W}atson {A}nalytics}.
\newblock
  \bibinfo{howpublished}{\url{http://www.ibm.com/analytics/watson-analytics}}.
\newblock


\bibitem[sup(2023)]%
        {superstore}
 \bibinfo{year}{2023}\natexlab{}.
\newblock \bibinfo{title}{Tableau {S}uperstore}.
\newblock
\newblock
\newblock
\shownote{CC-BY Dataset:
  \url{https://help.tableau.com/current/guides/get-started-tutorial/en-us/get-started-tutorial-connect.htm}}.


\bibitem[tho(2023)]%
        {thoughtspot}
 \bibinfo{year}{2023}\natexlab{}.
\newblock \bibinfo{title}{{T}hought{S}pot}.
\newblock \bibinfo{howpublished}{\url{http://www.thoughtspot.com}}.
\newblock


\bibitem[us-(2023)]%
        {us-crimes}
 \bibinfo{year}{2023}\natexlab{}.
\newblock \bibinfo{title}{{U.S.} {C}rimes}.
\newblock
\newblock
\newblock
\shownote{CC-BY Dataset:
  \url{https://www.kaggle.com/datasets/johnybhiduri/us-crime-data}}.


\bibitem[hou(2023)]%
        {housing}
 \bibinfo{year}{2023}\natexlab{}.
\newblock \bibinfo{title}{{U.S.} {H}ouse {L}istings}.
\newblock
\newblock
\newblock
\shownote{CC-BY Dataset:
  \url{https://www.kaggle.com/datasets/austinreese/usa-housing-listings}}.


\bibitem[Adelfio and Samet(2013)]%
        {adelfio:2013}
\bibfield{author}{\bibinfo{person}{Marco~D. Adelfio} {and}
  \bibinfo{person}{Hanan Samet}.} \bibinfo{year}{2013}\natexlab{}.
\newblock \showarticletitle{Schema Extraction for Tabular Data on the Web}.
\newblock \bibinfo{journal}{\emph{Proc. VLDB Endow.}} \bibinfo{volume}{6},
  \bibinfo{number}{6} (\bibinfo{date}{apr} \bibinfo{year}{2013}),
  \bibinfo{pages}{421–432}.
\newblock
\showISSN{2150-8097}
\urldef\tempurl%
\url{https://doi.org/10.14778/2536336.2536343}
\showDOI{\tempurl}


\bibitem[Ahlberg and Shneiderman(1994)]%
        {ahlberg1994visual}
\bibfield{author}{\bibinfo{person}{Christopher Ahlberg} {and}
  \bibinfo{person}{Ben Shneiderman}.} \bibinfo{year}{1994}\natexlab{}.
\newblock \showarticletitle{Visual information seeking: Tight coupling of
  dynamic query filters with starfield displays}. In
  \bibinfo{booktitle}{\emph{Proceedings of the SIGCHI conference on Human
  factors in computing systems}}. \bibinfo{pages}{313--317}.
\newblock


\bibitem[Airio et~al\mbox{.}(2004)]%
        {ciri}
\bibfield{author}{\bibinfo{person}{Eija Airio}, \bibinfo{person}{Kalervo
  J{\"a}rvelin}, \bibinfo{person}{Pirkko Saatsi}, \bibinfo{person}{Jaana
  Kek{\"a}l{\"a}inen}, {and} \bibinfo{person}{Sari Suomela}.}
  \bibinfo{year}{2004}\natexlab{}.
\newblock \bibinfo{booktitle}{\emph{CIRI - An Ontology-based Query Interface
  for Text Retrieval} (\bibinfo{edition}{1} ed.)}.
\newblock Number~20 in \bibinfo{series}{Publications of the Finnish Artificial
  Intelligence Society}. \bibinfo{publisher}{Finnish Artificial Intelligence
  Society}, \bibinfo{pages}{73--82}.
\newblock
\showISBNx{951-96735-55}


\bibitem[Bansal(2021)]%
        {bansal_2021}
\bibfield{author}{\bibinfo{person}{Shivam Bansal}.}
  \bibinfo{year}{2021}\natexlab{}.
\newblock \bibinfo{title}{Netflix movies and TV shows}.
\newblock
  \bibinfo{howpublished}{\url{https://www.kaggle.com/shivamb/netflix-shows}}.
\newblock
\newblock
\shownote{Accessed: 2023}.


\bibitem[Barr et~al\mbox{.}(2008)]%
        {barr2008linguistic}
\bibfield{author}{\bibinfo{person}{Cory Barr}, \bibinfo{person}{Rosie Jones},
  {and} \bibinfo{person}{Moira Regelson}.} \bibinfo{year}{2008}\natexlab{}.
\newblock \showarticletitle{The linguistic structure of English web-search
  queries}. In \bibinfo{booktitle}{\emph{Proceedings of the 2008 Conference on
  Empirical Methods in Natural Language Processing}}.
  \bibinfo{pages}{1021--1030}.
\newblock


\bibitem[Battle et~al\mbox{.}(2021)]%
        {battle2021exploring}
\bibfield{author}{\bibinfo{person}{Leilani Battle}, \bibinfo{person}{Danni
  Feng}, {and} \bibinfo{person}{Kelli Webber}.}
  \bibinfo{year}{2021}\natexlab{}.
\newblock \showarticletitle{Exploring Visualization Implementation Challenges
  Faced by D3 Users Online}.
\newblock \bibinfo{journal}{\emph{arXiv preprint arXiv:2108.02299}}
  (\bibinfo{year}{2021}).
\newblock


\bibitem[Bhagdev et~al\mbox{.}(2008)]%
        {bhagdev:2008}
\bibfield{author}{\bibinfo{person}{Ravish Bhagdev}, \bibinfo{person}{Sam
  Chapman}, \bibinfo{person}{Fabio Ciravegna}, \bibinfo{person}{Vitaveska
  Lanfranchi}, {and} \bibinfo{person}{Daniela Petrelli}.}
  \bibinfo{year}{2008}\natexlab{}.
\newblock \showarticletitle{Hybrid Search: Effectively Combining Keywords and
  Semantic Searches}. \bibinfo{pages}{554--568}.
\newblock
\showISBNx{978-3-540-68233-2}
\urldef\tempurl%
\url{https://doi.org/10.1007/978-3-540-68234-9_41}
\showDOI{\tempurl}


\bibitem[Bhaybhang(2022)]%
        {coffeesales}
\bibfield{author}{\bibinfo{person}{Arjun Bhaybhang}.}
  \bibinfo{year}{2022}\natexlab{}.
\newblock \bibinfo{title}{Coffee {C}hains {D}ataset}.
\newblock
  \bibinfo{howpublished}{\url{https://www.kaggle.com/datasets/arjunbhaybhang/coffee-chains-dataset}}.
\newblock
\newblock
\shownote{Accessed: 2023}.


\bibitem[Bostock et~al\mbox{.}(2011)]%
        {2011-d3}
\bibfield{author}{\bibinfo{person}{Michael Bostock}, \bibinfo{person}{Vadim
  Ogievetsky}, {and} \bibinfo{person}{Jeffrey Heer}.}
  \bibinfo{year}{2011}\natexlab{}.
\newblock \showarticletitle{D3: Data-Driven Documents}.
\newblock \bibinfo{journal}{\emph{IEEE Transactions on Visualization and
  Computer Graphics}} (\bibinfo{year}{2011}).
\newblock
\urldef\tempurl%
\url{http://idl.cs.washington.edu/papers/d3}
\showURL{%
\tempurl}


\bibitem[Buscaldi et~al\mbox{.}(2005)]%
        {Buscaldi2005AWQ}
\bibfield{author}{\bibinfo{person}{D. Buscaldi}, \bibinfo{person}{Paolo Rosso},
  {and} \bibinfo{person}{Emilio~Sanchis Arnal}.}
  \bibinfo{year}{2005}\natexlab{}.
\newblock \showarticletitle{A WordNet-based Query Expansion Method for
  Geographical Information Retrieval}. In \bibinfo{booktitle}{\emph{Conference
  and Labs of the Evaluation Forum}}.
\newblock


\bibitem[Cheng et~al\mbox{.}(2008)]%
        {falcons}
\bibfield{author}{\bibinfo{person}{Gong Cheng}, \bibinfo{person}{Weiyi Ge},
  {and} \bibinfo{person}{Yuzhong Qu}.} \bibinfo{year}{2008}\natexlab{}.
\newblock \showarticletitle{Falcons: Searching and Browsing Entities on the
  Semantic Web}. In \bibinfo{booktitle}{\emph{Proceedings of the 17th
  International Conference on World Wide Web}} (Beijing, China)
  \emph{(\bibinfo{series}{WWW '08})}. \bibinfo{publisher}{Association for
  Computing Machinery}, \bibinfo{address}{New York, NY, USA},
  \bibinfo{pages}{1101–1102}.
\newblock
\showISBNx{9781605580852}
\urldef\tempurl%
\url{https://doi.org/10.1145/1367497.1367676}
\showDOI{\tempurl}


\bibitem[Christodoulakis et~al\mbox{.}(2020)]%
        {pytheus}
\bibfield{author}{\bibinfo{person}{Christina Christodoulakis},
  \bibinfo{person}{Eric~B. Munson}, \bibinfo{person}{Moshe Gabel},
  \bibinfo{person}{Angela~Demke Brown}, {and} \bibinfo{person}{Ren\'{e}e~J.
  Miller}.} \bibinfo{year}{2020}\natexlab{}.
\newblock \showarticletitle{Pytheas: Pattern-Based Table Discovery in CSV
  Files}.
\newblock \bibinfo{journal}{\emph{Proc. VLDB Endow.}} \bibinfo{volume}{13},
  \bibinfo{number}{12} (\bibinfo{date}{jul} \bibinfo{year}{2020}),
  \bibinfo{pages}{2075–2089}.
\newblock
\showISSN{2150-8097}
\urldef\tempurl%
\url{https://doi.org/10.14778/3407790.3407810}
\showDOI{\tempurl}


\bibitem[Chu et~al\mbox{.}(2016)]%
        {datacleaning:survey}
\bibfield{author}{\bibinfo{person}{Xu Chu}, \bibinfo{person}{Ihab~F. Ilyas},
  \bibinfo{person}{Sanjay Krishnan}, {and} \bibinfo{person}{Jiannan Wang}.}
  \bibinfo{year}{2016}\natexlab{}.
\newblock \showarticletitle{Data Cleaning: Overview and Emerging Challenges}.
  In \bibinfo{booktitle}{\emph{Proceedings of the 2016 International Conference
  on Management of Data}} (San Francisco, California, USA)
  \emph{(\bibinfo{series}{SIGMOD '16})}. \bibinfo{publisher}{Association for
  Computing Machinery}, \bibinfo{address}{New York, NY, USA},
  \bibinfo{pages}{2201–2206}.
\newblock
\showISBNx{9781450335317}
\urldef\tempurl%
\url{https://doi.org/10.1145/2882903.2912574}
\showDOI{\tempurl}


\bibitem[Chu-Carroll et~al\mbox{.}(2006)]%
        {chu-carroll:2006}
\bibfield{author}{\bibinfo{person}{Jennifer Chu-Carroll}, \bibinfo{person}{John
  Prager}, \bibinfo{person}{Krzysztof Czuba}, \bibinfo{person}{David Ferrucci},
  {and} \bibinfo{person}{Pablo Duboue}.} \bibinfo{year}{2006}\natexlab{}.
\newblock \showarticletitle{Semantic Search via XML Fragments: A High-Precision
  Approach to IR}. In \bibinfo{booktitle}{\emph{Proceedings of the 29th Annual
  International ACM SIGIR Conference on Research and Development in Information
  Retrieval}} (Seattle, Washington, USA) \emph{(\bibinfo{series}{SIGIR '06})}.
  \bibinfo{publisher}{Association for Computing Machinery},
  \bibinfo{address}{New York, NY, USA}, \bibinfo{pages}{445–452}.
\newblock
\showISBNx{1595933697}
\urldef\tempurl%
\url{https://doi.org/10.1145/1148170.1148247}
\showDOI{\tempurl}


\bibitem[Cimiano et~al\mbox{.}(2008)]%
        {cimiano:2008}
\bibfield{author}{\bibinfo{person}{Philipp Cimiano}, \bibinfo{person}{Peter
  Haase}, \bibinfo{person}{J\"{o}rg Heizmann}, \bibinfo{person}{Matthias
  Mantel}, {and} \bibinfo{person}{Rudi Studer}.}
  \bibinfo{year}{2008}\natexlab{}.
\newblock \showarticletitle{Towards Portable Natural Language Interfaces to
  Knowledge Bases - The Case of the ORAKEL System}.
\newblock \bibinfo{journal}{\emph{Data Knowl. Eng.}} \bibinfo{volume}{65},
  \bibinfo{number}{2} (\bibinfo{date}{may} \bibinfo{year}{2008}),
  \bibinfo{pages}{325–354}.
\newblock
\showISSN{0169-023X}
\urldef\tempurl%
\url{https://doi.org/10.1016/j.datak.2007.10.007}
\showDOI{\tempurl}


\bibitem[Cocke(1969)]%
        {cocke1969programming}
\bibfield{author}{\bibinfo{person}{John Cocke}.}
  \bibinfo{year}{1969}\natexlab{}.
\newblock \bibinfo{booktitle}{\emph{Programming languages and their compilers:
  Preliminary notes}}.
\newblock \bibinfo{publisher}{New York University}.
\newblock


\bibitem[Constantino(2023)]%
        {tableau_server_search}
\bibfield{author}{\bibinfo{person}{Joe Constantino}.}
  \bibinfo{year}{2023}\natexlab{}.
\newblock \bibinfo{title}{The Evolution of Tableau Search and Best Practices
  for Finding Relevant Content}.
\newblock
  \bibinfo{howpublished}{\url{https://www.tableau.com/blog/evolution-tableau-search-and-best-practices-finding-relevant-content}}.
\newblock


\bibitem[Corby et~al\mbox{.}(2004)]%
        {corby:2004}
\bibfield{author}{\bibinfo{person}{Olivier Corby}, \bibinfo{person}{Rose
  Dieng-Kuntz}, {and} \bibinfo{person}{Catherine Faron-Zucker}.}
  \bibinfo{year}{2004}\natexlab{}.
\newblock \showarticletitle{Querying the Semantic Web with the CORESE search
  engine}.
\newblock \bibinfo{journal}{\emph{Proceedings of the 16th European Conference
  on Artificial Intelligence (ECAI'2004)}}, \bibinfo{pages}{705--709}.
\newblock


\bibitem[Crook et~al\mbox{.}(2018)]%
        {crook:2018}
\bibfield{author}{\bibinfo{person}{Paul~A. Crook}, \bibinfo{person}{Alex
  Marin}, \bibinfo{person}{Vipul Agarwal}, \bibinfo{person}{Samantha Anderson},
  \bibinfo{person}{Ohyoung Jang}, \bibinfo{person}{Aliasgar Lanewala},
  \bibinfo{person}{Karthik Tangirala}, {and} \bibinfo{person}{Imed Zitouni}.}
  \bibinfo{year}{2018}\natexlab{}.
\newblock \showarticletitle{Conversational Semantic Search: Looking Beyond Web
  Search,\& and Dialog Systems}. In \bibinfo{booktitle}{\emph{Proceedings of
  the Eleventh ACM International Conference on Web Search and Data Mining}}
  (Marina Del Rey, CA, USA) \emph{(\bibinfo{series}{WSDM '18})}.
  \bibinfo{publisher}{Association for Computing Machinery},
  \bibinfo{address}{New York, NY, USA}, \bibinfo{pages}{763–766}.
\newblock
\showISBNx{9781450355810}
\urldef\tempurl%
\url{https://doi.org/10.1145/3159652.3160590}
\showDOI{\tempurl}


\bibitem[Cui et~al\mbox{.}(2019)]%
        {cui2019datasite}
\bibfield{author}{\bibinfo{person}{Zhe Cui}, \bibinfo{person}{Sriram~Karthik
  Badam}, \bibinfo{person}{M~Adil Yal{\c{c}}in}, {and} \bibinfo{person}{Niklas
  Elmqvist}.} \bibinfo{year}{2019}\natexlab{}.
\newblock \showarticletitle{Datasite: Proactive visual data exploration with
  computation of insight-based recommendations}.
\newblock \bibinfo{journal}{\emph{Information Visualization}}
  \bibinfo{volume}{18}, \bibinfo{number}{2} (\bibinfo{year}{2019}),
  \bibinfo{pages}{251--267}.
\newblock


\bibitem[Damljanovic et~al\mbox{.}(2010)]%
        {Damljanovic2010NaturalLI}
\bibfield{author}{\bibinfo{person}{Danica Damljanovic}, \bibinfo{person}{Milan
  Agatonovic}, {and} \bibinfo{person}{Hamish Cunningham}.}
  \bibinfo{year}{2010}\natexlab{}.
\newblock \showarticletitle{Natural Language Interfaces to Ontologies:
  Combining Syntactic Analysis and Ontology-Based Lookup through the User
  Interaction}. In \bibinfo{booktitle}{\emph{Extended Semantic Web
  Conference}}.
\newblock


\bibitem[Demiralp et~al\mbox{.}(2017)]%
        {demiralp2017foresight}
\bibfield{author}{\bibinfo{person}{{\c{C}}a{\u{g}}atay Demiralp},
  \bibinfo{person}{Peter~J Haas}, \bibinfo{person}{Srinivasan Parthasarathy},
  {and} \bibinfo{person}{Tejaswini Pedapati}.} \bibinfo{year}{2017}\natexlab{}.
\newblock \showarticletitle{Foresight: Recommending visual insights}.
\newblock \bibinfo{journal}{\emph{arXiv preprint arXiv:1707.03877}}
  (\bibinfo{year}{2017}).
\newblock


\bibitem[Dhamdhere et~al\mbox{.}(2017)]%
        {analyza}
\bibfield{author}{\bibinfo{person}{Kedar Dhamdhere}, \bibinfo{person}{Kevin~S.
  McCurley}, \bibinfo{person}{Ralfi Nahmias}, \bibinfo{person}{Mukund
  Sundararajan}, {and} \bibinfo{person}{Qiqi Yan}.}
  \bibinfo{year}{2017}\natexlab{}.
\newblock \showarticletitle{Analyza: Exploring Data with Conversation}. In
  \bibinfo{booktitle}{\emph{Proceedings of the 22nd International Conference on
  Intelligent User Interfaces}} \emph{(\bibinfo{series}{IUI 2017})}.
  \bibinfo{pages}{493--504}.
\newblock


\bibitem[Ding et~al\mbox{.}(2005)]%
        {ding:2005}
\bibfield{author}{\bibinfo{person}{Li Ding}, \bibinfo{person}{Tim Finin},
  \bibinfo{person}{Anupam Joshi}, \bibinfo{person}{Yun Peng},
  \bibinfo{person}{Rong Pan}, {and} \bibinfo{person}{Pavan Reddivari}.}
  \bibinfo{year}{2005}\natexlab{}.
\newblock \showarticletitle{Search on the Semantic Web}.
\newblock \bibinfo{journal}{\emph{Computer}}  \bibinfo{volume}{38}
  (\bibinfo{date}{11} \bibinfo{year}{2005}), \bibinfo{pages}{62 -- 69}.
\newblock
\urldef\tempurl%
\url{https://doi.org/10.1109/MC.2005.350}
\showDOI{\tempurl}


\bibitem[Fasciano and Lapalme(1996)]%
        {fasciano-lapalme-1996-postgraphe}
\bibfield{author}{\bibinfo{person}{Massimo Fasciano} {and} \bibinfo{person}{Guy
  Lapalme}.} \bibinfo{year}{1996}\natexlab{}.
\newblock \showarticletitle{{P}ost{G}raphe: A System for the Generation of
  Statistical Graphics and Text}. In \bibinfo{booktitle}{\emph{Eighth
  International Natural Language Generation Workshop}}.
\newblock
\urldef\tempurl%
\url{https://aclanthology.org/W96-0406}
\showURL{%
\tempurl}


\bibitem[Fellbaum(1998)]%
        {wordnet}
\bibfield{author}{\bibinfo{person}{Christiane Fellbaum}.}
  \bibinfo{year}{1998}\natexlab{}.
\newblock \bibinfo{booktitle}{\emph{WordNet: An Electronic Lexical Database}}.
\newblock \bibinfo{publisher}{Bradford Books}.
\newblock


\bibitem[Fernandez et~al\mbox{.}(2008)]%
        {fernandez:2008}
\bibfield{author}{\bibinfo{person}{Miriam Fernandez}, \bibinfo{person}{Vanessa
  Lopez}, \bibinfo{person}{Marta Sabou}, \bibinfo{person}{Victoria Uren},
  \bibinfo{person}{David Vallet}, \bibinfo{person}{Enrico Motta}, {and}
  \bibinfo{person}{Pablo Castells}.} \bibinfo{year}{2008}\natexlab{}.
\newblock \showarticletitle{Semantic Search Meets the Web}. In
  \bibinfo{booktitle}{\emph{2008 IEEE International Conference on Semantic
  Computing}}. \bibinfo{pages}{253--260}.
\newblock
\urldef\tempurl%
\url{https://doi.org/10.1109/ICSC.2008.52}
\showDOI{\tempurl}


\bibitem[Finin et~al\mbox{.}(2005)]%
        {swoogle}
\bibfield{author}{\bibinfo{person}{Tim Finin}, \bibinfo{person}{Li Ding},
  \bibinfo{person}{Rong Pan}, \bibinfo{person}{Anupam Joshi},
  \bibinfo{person}{Pranam Kolari}, \bibinfo{person}{Akshay Java}, {and}
  \bibinfo{person}{Yun Peng}.} \bibinfo{year}{2005}\natexlab{}.
\newblock \showarticletitle{Swoogle: Searching for Knowledge on the Semantic
  Web}. In \bibinfo{booktitle}{\emph{Proceedings of the 20th National
  Conference on Artificial Intelligence - Volume 4}} (Pittsburgh, Pennsylvania)
  \emph{(\bibinfo{series}{AAAI'05})}. \bibinfo{publisher}{AAAI Press},
  \bibinfo{pages}{1682–1683}.
\newblock
\showISBNx{157735236x}


\bibitem[Fr{\"a}nti et~al\mbox{.}(2005)]%
        {franti2005mobile}
\bibfield{author}{\bibinfo{person}{Pasi Fr{\"a}nti}, \bibinfo{person}{Jaakko
  Sauvola}, {and} \bibinfo{person}{Hannu T{\"o}rm{\"a}nen}.}
  \bibinfo{year}{2005}\natexlab{}.
\newblock \showarticletitle{Mobile information retrieval with local intent
  analysis}. In \bibinfo{booktitle}{\emph{Mobile and Ubiquitous Information
  Access}}. Springer, \bibinfo{pages}{179--192}.
\newblock


\bibitem[Freedman et~al\mbox{.}(2007)]%
        {freedman2007statistics}
\bibfield{author}{\bibinfo{person}{David Freedman}, \bibinfo{person}{Robert
  Pisani}, {and} \bibinfo{person}{Roger Purves}.}
  \bibinfo{year}{2007}\natexlab{}.
\newblock \showarticletitle{Statistics (international student edition)}.
\newblock \bibinfo{journal}{\emph{Pisani, R. Purves, 4th edn. WW Norton \&
  Company, New York}} (\bibinfo{year}{2007}).
\newblock


\bibitem[Frieder et~al\mbox{.}(2023)]%
        {frieder2023mathematical}
\bibfield{author}{\bibinfo{person}{Simon Frieder}, \bibinfo{person}{Luca
  Pinchetti}, \bibinfo{person}{Ryan-Rhys Griffiths}, \bibinfo{person}{Tommaso
  Salvatori}, \bibinfo{person}{Thomas Lukasiewicz},
  \bibinfo{person}{Philipp~Christian Petersen}, \bibinfo{person}{Alexis
  Chevalier}, {and} \bibinfo{person}{Julius Berner}.}
  \bibinfo{year}{2023}\natexlab{}.
\newblock \bibinfo{title}{Mathematical Capabilities of ChatGPT}.
\newblock
\newblock
\showeprint[arxiv]{2301.13867}~[cs.LG]


\bibitem[Galhotra and Khurana(2020)]%
        {galhotra:2020}
\bibfield{author}{\bibinfo{person}{Sainyam Galhotra} {and}
  \bibinfo{person}{Udayan Khurana}.} \bibinfo{year}{2020}\natexlab{}.
\newblock \showarticletitle{Semantic Search over Structured Data}. In
  \bibinfo{booktitle}{\emph{Proceedings of the 29th ACM International
  Conference on Information \& Knowledge Management}} (Virtual Event, Ireland)
  \emph{(\bibinfo{series}{CIKM '20})}. \bibinfo{publisher}{Association for
  Computing Machinery}, \bibinfo{address}{New York, NY, USA},
  \bibinfo{pages}{3381–3384}.
\newblock
\showISBNx{9781450368599}
\urldef\tempurl%
\url{https://doi.org/10.1145/3340531.3417426}
\showDOI{\tempurl}


\bibitem[Gao et~al\mbox{.}(2015)]%
        {datatone}
\bibfield{author}{\bibinfo{person}{Tong Gao}, \bibinfo{person}{Mira Dontcheva},
  \bibinfo{person}{Eytan Adar}, \bibinfo{person}{Zhicheng Liu}, {and}
  \bibinfo{person}{Karrie~G. Karahalios}.} \bibinfo{year}{2015}\natexlab{}.
\newblock \showarticletitle{DataTone: Managing Ambiguity in Natural Language
  Interfaces for Data Visualization}. In \bibinfo{booktitle}{\emph{Proceedings
  of the 28th Annual ACM Symposium on User Interface Software Technology}}
  \emph{(\bibinfo{series}{UIST 2015})}. \bibinfo{publisher}{ACM},
  \bibinfo{address}{New York, NY, USA}, \bibinfo{pages}{489--500}.
\newblock
\showISBNx{978-1-4503-3779-3}


\bibitem[Gruber(1993)]%
        {grubar:1993}
\bibfield{author}{\bibinfo{person}{Thomas~R. Gruber}.}
  \bibinfo{year}{1993}\natexlab{}.
\newblock \showarticletitle{A translation approach to portable ontology
  specifications}.
\newblock \bibinfo{journal}{\emph{Knowledge Acquisition}} \bibinfo{volume}{5},
  \bibinfo{number}{2} (\bibinfo{year}{1993}), \bibinfo{pages}{199--220}.
\newblock
\showISSN{1042-8143}
\urldef\tempurl%
\url{https://doi.org/10.1006/knac.1993.1008}
\showDOI{\tempurl}


\bibitem[Guha et~al\mbox{.}(2003)]%
        {Guha2003SemanticS}
\bibfield{author}{\bibinfo{person}{Ramanathan~V. Guha}, \bibinfo{person}{Rob
  McCool}, {and} \bibinfo{person}{Eric Miller}.}
  \bibinfo{year}{2003}\natexlab{}.
\newblock \showarticletitle{Semantic search}. In \bibinfo{booktitle}{\emph{The
  Web Conference}}.
\newblock


\bibitem[Hammiche et~al\mbox{.}(2004)]%
        {hammiche:2004}
\bibfield{author}{\bibinfo{person}{Samira Hammiche}, \bibinfo{person}{Salima
  Benbernou}, \bibinfo{person}{Mohand-Sa\"{\i}d Hacid}, {and}
  \bibinfo{person}{Athena Vakali}.} \bibinfo{year}{2004}\natexlab{}.
\newblock \showarticletitle{Semantic Retrieval of Multimedia Data}. In
  \bibinfo{booktitle}{\emph{Proceedings of the 2nd ACM International Workshop
  on Multimedia Databases}} (Washington, DC, USA) \emph{(\bibinfo{series}{MMDB
  '04})}. \bibinfo{publisher}{Association for Computing Machinery},
  \bibinfo{address}{New York, NY, USA}, \bibinfo{pages}{36–44}.
\newblock
\showISBNx{1581139756}
\urldef\tempurl%
\url{https://doi.org/10.1145/1032604.1032612}
\showDOI{\tempurl}


\bibitem[Harper and Agrawala(2014)]%
        {d3deconstruction}
\bibfield{author}{\bibinfo{person}{Jonathan Harper} {and}
  \bibinfo{person}{Maneesh Agrawala}.} \bibinfo{year}{2014}\natexlab{}.
\newblock \showarticletitle{Deconstructing and restyling D3 visualizations}.
\newblock \bibinfo{journal}{\emph{UIST 2014 - Proceedings of the 27th Annual
  ACM Symposium on User Interface Software and Technology}} (\bibinfo{date}{10}
  \bibinfo{year}{2014}), \bibinfo{pages}{253--262}.
\newblock
\urldef\tempurl%
\url{https://doi.org/10.1145/2642918.2647411}
\showDOI{\tempurl}


\bibitem[Harth et~al\mbox{.}(2007)]%
        {Harth2007SWSEAB}
\bibfield{author}{\bibinfo{person}{A. Harth}, \bibinfo{person}{Aidan Hogan},
  \bibinfo{person}{Renaud Delbru}, \bibinfo{person}{J{\"u}rgen Umbrich},
  \bibinfo{person}{Se{\'a}n O'Riain}, {and} \bibinfo{person}{Stefan Decker}.}
  \bibinfo{year}{2007}\natexlab{}.
\newblock \showarticletitle{SWSE: Answers Before Links!}. In
  \bibinfo{booktitle}{\emph{Semantic Web Challenge}}.
\newblock


\bibitem[Hearst(2009)]%
        {hearst2009search}
\bibfield{author}{\bibinfo{person}{Marti Hearst}.}
  \bibinfo{year}{2009}\natexlab{}.
\newblock \bibinfo{booktitle}{\emph{Search user interfaces}}.
\newblock \bibinfo{publisher}{Cambridge university press}.
\newblock


\bibitem[Heflin et~al\mbox{.}(2003)]%
        {Heflin2003SHOEAB}
\bibfield{author}{\bibinfo{person}{Jeff Heflin}, \bibinfo{person}{James~A.
  Hendler}, {and} \bibinfo{person}{Sean Luke}.}
  \bibinfo{year}{2003}\natexlab{}.
\newblock \showarticletitle{SHOE: A Blueprint for the Semantic Web}. In
  \bibinfo{booktitle}{\emph{Spinning the Semantic Web}}.
\newblock


\bibitem[Hoque and Agrawala(2019)]%
        {hoque2019searching}
\bibfield{author}{\bibinfo{person}{Enamul Hoque} {and} \bibinfo{person}{Maneesh
  Agrawala}.} \bibinfo{year}{2019}\natexlab{}.
\newblock \showarticletitle{Searching the visual style and structure of {D3}
  visualizations}.
\newblock \bibinfo{journal}{\emph{IEEE Transactions on Visualization and
  Computer Graphics}} \bibinfo{volume}{26}, \bibinfo{number}{1}
  (\bibinfo{year}{2019}), \bibinfo{pages}{1236--1245}.
\newblock


\bibitem[Hoque et~al\mbox{.}(2017)]%
        {hoque2017applying}
\bibfield{author}{\bibinfo{person}{Enamul Hoque}, \bibinfo{person}{Vidya
  Setlur}, \bibinfo{person}{Melanie Tory}, {and} \bibinfo{person}{Isaac
  Dykeman}.} \bibinfo{year}{2017}\natexlab{}.
\newblock \showarticletitle{Applying pragmatics principles for interaction with
  visual analytics}.
\newblock \bibinfo{journal}{\emph{IEEE Transactions on Visualization and
  Computer Graphics}} \bibinfo{volume}{24}, \bibinfo{number}{1}
  (\bibinfo{year}{2017}), \bibinfo{pages}{309--318}.
\newblock


\bibitem[Jing and Baluja(2008)]%
        {visualrank}
\bibfield{author}{\bibinfo{person}{Yushi Jing} {and} \bibinfo{person}{Shumeet
  Baluja}.} \bibinfo{year}{2008}\natexlab{}.
\newblock \showarticletitle{VisualRank: Applying PageRank to Large-Scale Image
  Search}.
\newblock \bibinfo{journal}{\emph{IEEE Transactions on Pattern Analysis and
  Machine Intelligence}} \bibinfo{volume}{30}, \bibinfo{number}{11}
  (\bibinfo{year}{2008}), \bibinfo{pages}{1877--1890}.
\newblock
\urldef\tempurl%
\url{https://doi.org/10.1109/TPAMI.2008.121}
\showDOI{\tempurl}


\bibitem[Jupyter(2023)]%
        {jupyter}
\bibfield{author}{\bibinfo{person}{Jupyter}.} \bibinfo{year}{2023}\natexlab{}.
\newblock \bibinfo{title}{Jupyter}.
\newblock \bibinfo{howpublished}{\url{https://jupyter.org/}}.
\newblock


\bibitem[Kandel et~al\mbox{.}(2011)]%
        {wrangler}
\bibfield{author}{\bibinfo{person}{Sean Kandel}, \bibinfo{person}{Andreas
  Paepcke}, \bibinfo{person}{Joseph Hellerstein}, {and}
  \bibinfo{person}{Jeffrey Heer}.} \bibinfo{year}{2011}\natexlab{}.
\newblock \showarticletitle{Wrangler: Interactive Visual Specification of Data
  Transformation Scripts}. In \bibinfo{booktitle}{\emph{Proceedings of the
  SIGCHI Conference on Human Factors in Computing Systems}} (Vancouver, BC,
  Canada) \emph{(\bibinfo{series}{CHI '11})}. \bibinfo{publisher}{Association
  for Computing Machinery}, \bibinfo{address}{New York, NY, USA},
  \bibinfo{pages}{3363–3372}.
\newblock
\showISBNx{9781450302289}
\urldef\tempurl%
\url{https://doi.org/10.1145/1978942.1979444}
\showDOI{\tempurl}


\bibitem[Kasami(1966)]%
        {kasami1966efficient}
\bibfield{author}{\bibinfo{person}{Tadao Kasami}.}
  \bibinfo{year}{1966}\natexlab{}.
\newblock \showarticletitle{An efficient recognition and syntax-analysis
  algorithm for context-free languages}.
\newblock \bibinfo{journal}{\emph{Coordinated Science Laboratory Report no.
  R-257}} (\bibinfo{year}{1966}).
\newblock


\bibitem[Kasneci et~al\mbox{.}(2008)]%
        {Kasneci2008NAGAHS}
\bibfield{author}{\bibinfo{person}{Gjergji Kasneci}, \bibinfo{person}{Fabian~M.
  Suchanek}, \bibinfo{person}{Georgiana Ifrim}, \bibinfo{person}{Shady
  Elbassuoni}, \bibinfo{person}{Maya Ramanath}, {and} \bibinfo{person}{Gerhard
  Weikum}.} \bibinfo{year}{2008}\natexlab{}.
\newblock \showarticletitle{NAGA: harvesting, searching and ranking knowledge}.
  In \bibinfo{booktitle}{\emph{SIGMOD Conference}}.
\newblock


\bibitem[Kaufmann et~al\mbox{.}(2006)]%
        {kaufmann:2006}
\bibfield{author}{\bibinfo{person}{Esther Kaufmann}, \bibinfo{person}{Abraham
  Bernstein}, {and} \bibinfo{person}{Renato Zumstein}.}
  \bibinfo{year}{2006}\natexlab{}.
\newblock \showarticletitle{Querix: A natural language interface to query
  ontologies based on clarification dialogs}.
\newblock  (\bibinfo{date}{01} \bibinfo{year}{2006}).
\newblock


\bibitem[Kim et~al\mbox{.}(2021)]%
        {kim:2021}
\bibfield{author}{\bibinfo{person}{Dae~Hyun Kim}, \bibinfo{person}{Vidya
  Setlur}, {and} \bibinfo{person}{Maneesh Agrawala}.}
  \bibinfo{year}{2021}\natexlab{}.
\newblock \showarticletitle{Towards Understanding How Readers Integrate Charts
  and Captions: A Case Study with Line Charts}. In
  \bibinfo{booktitle}{\emph{Proceedings of the 2021 CHI Conference on Human
  Factors in Computing Systems}} (Yokohama, Japan) \emph{(\bibinfo{series}{CHI
  '21})}. \bibinfo{publisher}{Association for Computing Machinery},
  \bibinfo{address}{New York, NY, USA}, Article \bibinfo{articleno}{610},
  \bibinfo{numpages}{11}~pages.
\newblock
\showISBNx{9781450380966}
\urldef\tempurl%
\url{https://doi.org/10.1145/3411764.3445443}
\showDOI{\tempurl}


\bibitem[Klein and Manning(2003)]%
        {klein-manning-2003-accurate}
\bibfield{author}{\bibinfo{person}{Dan Klein} {and}
  \bibinfo{person}{Christopher~D. Manning}.} \bibinfo{year}{2003}\natexlab{}.
\newblock \showarticletitle{Accurate Unlexicalized Parsing}. In
  \bibinfo{booktitle}{\emph{Proceedings of the 41st Annual Meeting of the
  Association for Computational Linguistics}}. \bibinfo{publisher}{Association
  for Computational Linguistics}, \bibinfo{address}{Sapporo, Japan},
  \bibinfo{pages}{423--430}.
\newblock
\urldef\tempurl%
\url{https://doi.org/10.3115/1075096.1075150}
\showDOI{\tempurl}


\bibitem[Kolomiyets and Moens(2011)]%
        {KOLOMIYETS20115412}
\bibfield{author}{\bibinfo{person}{Oleksandr Kolomiyets} {and}
  \bibinfo{person}{Marie-Francine Moens}.} \bibinfo{year}{2011}\natexlab{}.
\newblock \showarticletitle{A survey on question answering technology from an
  information retrieval perspective}.
\newblock \bibinfo{journal}{\emph{Information Sciences}} \bibinfo{volume}{181},
  \bibinfo{number}{24} (\bibinfo{year}{2011}), \bibinfo{pages}{5412--5434}.
\newblock
\showISSN{0020-0255}
\urldef\tempurl%
\url{https://doi.org/10.1016/j.ins.2011.07.047}
\showDOI{\tempurl}


\bibitem[Lei et~al\mbox{.}(2006)]%
        {Lei2006SemSearchAS}
\bibfield{author}{\bibinfo{person}{Yuangui Lei}, \bibinfo{person}{Victoria~S.
  Uren}, {and} \bibinfo{person}{Enrico Motta}.}
  \bibinfo{year}{2006}\natexlab{}.
\newblock \showarticletitle{SemSearch: A Search Engine for the Semantic Web}.
  In \bibinfo{booktitle}{\emph{International Conference Knowledge Engineering
  and Knowledge Management}}.
\newblock


\bibitem[L{\'o}pez et~al\mbox{.}(2005)]%
        {Lpez2005AquaLogAO}
\bibfield{author}{\bibinfo{person}{V. L{\'o}pez}, \bibinfo{person}{Michele
  Pasin}, {and} \bibinfo{person}{Enrico Motta}.}
  \bibinfo{year}{2005}\natexlab{}.
\newblock \showarticletitle{AquaLog: An Ontology-Portable Question Answering
  System for the Semantic Web}. In \bibinfo{booktitle}{\emph{Extended Semantic
  Web Conference}}.
\newblock


\bibitem[Lopez et~al\mbox{.}(2006)]%
        {lopez:2006}
\bibfield{author}{\bibinfo{person}{Vanessa Lopez}, \bibinfo{person}{Marta
  Sabou}, {and} \bibinfo{person}{Enrico Motta}.}
  \bibinfo{year}{2006}\natexlab{}.
\newblock \showarticletitle{PowerMap: Mapping the Real Semantic Web on the
  Fly}. In \bibinfo{booktitle}{\emph{Proceedings of the 5th International
  Conference on The Semantic Web}} (Athens, GA)
  \emph{(\bibinfo{series}{ISWC'06})}. \bibinfo{publisher}{Springer-Verlag},
  \bibinfo{address}{Berlin, Heidelberg}, \bibinfo{pages}{414–427}.
\newblock
\showISBNx{3540490299}
\urldef\tempurl%
\url{https://doi.org/10.1007/11926078_30}
\showDOI{\tempurl}


\bibitem[Mackinlay et~al\mbox{.}(2007)]%
        {mackinlay2007show}
\bibfield{author}{\bibinfo{person}{Jock Mackinlay}, \bibinfo{person}{Pat
  Hanrahan}, {and} \bibinfo{person}{Chris Stolte}.}
  \bibinfo{year}{2007}\natexlab{}.
\newblock \showarticletitle{Show me: Automatic presentation for visual
  analysis}.
\newblock \bibinfo{journal}{\emph{IEEE Transactions on Visualization and
  Computer Graphics}} \bibinfo{volume}{13}, \bibinfo{number}{6}
  (\bibinfo{year}{2007}), \bibinfo{pages}{1137--1144}.
\newblock


\bibitem[Manning et~al\mbox{.}(2008)]%
        {manning2008introduction}
\bibfield{author}{\bibinfo{person}{Christopher~D. Manning},
  \bibinfo{person}{Prabhakar Raghavan}, {and} \bibinfo{person}{Hinrich
  Schütze}.} \bibinfo{year}{2008}\natexlab{}.
\newblock \bibinfo{booktitle}{\emph{Introduction to Information Retrieval}}.
\newblock \bibinfo{publisher}{Cambridge University Press},
  \bibinfo{address}{Cambridge, UK}.
\newblock
\showISBNx{978-0-521-86571-5}
\urldef\tempurl%
\url{http://nlp.stanford.edu/IR-book/information-retrieval-book.html}
\showURL{%
\tempurl}


\bibitem[Marchionini(2006)]%
        {marchionini2006exploratory}
\bibfield{author}{\bibinfo{person}{Gary Marchionini}.}
  \bibinfo{year}{2006}\natexlab{}.
\newblock \showarticletitle{Exploratory search: from finding to understanding}.
\newblock \bibinfo{journal}{\emph{Commun. ACM}} \bibinfo{volume}{49},
  \bibinfo{number}{4} (\bibinfo{year}{2006}), \bibinfo{pages}{41--46}.
\newblock


\bibitem[Merriam-Webster(2023)]%
        {mw:olio}
\bibfield{author}{\bibinfo{person}{Merriam-Webster}.}
  \bibinfo{year}{2023}\natexlab{}.
\newblock \showarticletitle{Olio}. In \bibinfo{booktitle}{\emph{Merriam-Webster
  {D}ictionary}}.
\newblock
\urldef\tempurl%
\url{https://www.merriam-webster.com/dictionary/olio}
\showURL{%
\tempurl}


\bibitem[{M}etadata(2023)]%
        {Observable:metadata}
\bibfield{author}{\bibinfo{person}{Observable {M}etadata}.}
  \bibinfo{year}{2023}\natexlab{}.
\newblock \bibinfo{title}{Searching on {O}bservable}.
\newblock
  \bibinfo{howpublished}{\url{https://observablehq.com/@observablehq/searching-on-observable\#attributes}}.
\newblock


\bibitem[Meyer et~al\mbox{.}(2022)]%
        {meyer:2022}
\bibfield{author}{\bibinfo{person}{Selina Meyer}, \bibinfo{person}{David
  Elsweiler}, \bibinfo{person}{Bernd Ludwig}, \bibinfo{person}{Marcos
  Fernandez-Pichel}, {and} \bibinfo{person}{David~E. Losada}.}
  \bibinfo{year}{2022}\natexlab{}.
\newblock \showarticletitle{Do We Still Need Human Assessors? Prompt-Based
  GPT-3 User Simulation in Conversational AI}. In
  \bibinfo{booktitle}{\emph{Proceedings of the 4th Conference on Conversational
  User Interfaces}} (Glasgow, United Kingdom) \emph{(\bibinfo{series}{CUI
  '22})}. \bibinfo{publisher}{Association for Computing Machinery},
  \bibinfo{address}{New York, NY, USA}, Article \bibinfo{articleno}{8},
  \bibinfo{numpages}{6}~pages.
\newblock
\showISBNx{9781450397391}
\urldef\tempurl%
\url{https://doi.org/10.1145/3543829.3544529}
\showDOI{\tempurl}


\bibitem[Microsoft(2023)]%
        {powerbi}
\bibfield{author}{\bibinfo{person}{Microsoft}.}
  \bibinfo{year}{2023}\natexlab{}.
\newblock \bibinfo{title}{Microsoft PowerBI}.
\newblock \bibinfo{howpublished}{\url{https://powerbi.microsoft.com/}}.
\newblock


\bibitem[Mikolov et~al\mbox{.}(2013)]%
        {word2vec}
\bibfield{author}{\bibinfo{person}{Tom{\'{a}}s Mikolov}, \bibinfo{person}{Kai
  Chen}, \bibinfo{person}{Greg Corrado}, {and} \bibinfo{person}{Jeffrey Dean}.}
  \bibinfo{year}{2013}\natexlab{}.
\newblock \showarticletitle{Efficient Estimation of Word Representations in
  Vector Space}. In \bibinfo{booktitle}{\emph{1st International Conference on
  Learning Representations, {ICLR} 2013, Scottsdale, Arizona, USA, May 2-4,
  2013, Workshop Track Proceedings}}, \bibfield{editor}{\bibinfo{person}{Yoshua
  Bengio} {and} \bibinfo{person}{Yann LeCun}} (Eds.).
\newblock
\urldef\tempurl%
\url{http://arxiv.org/abs/1301.3781}
\showURL{%
\tempurl}


\bibitem[Mittal et~al\mbox{.}(1995)]%
        {mittal:1995}
\bibfield{author}{\bibinfo{person}{Vibhu Mittal}, \bibinfo{person}{Steven
  Roth}, \bibinfo{person}{Johanna Moore}, \bibinfo{person}{Joe Mattis}, {and}
  \bibinfo{person}{Giuseppe Carenini}.} \bibinfo{year}{1995}\natexlab{}.
\newblock \showarticletitle{Generating Explanatory Captions for Information
  Graphics}.
\newblock \bibinfo{journal}{\emph{Proceeedings of the International Joint
  Conference on Artificial Intelligence}}, \bibinfo{pages}{1276--1283}.
\newblock


\bibitem[Moldovan and Mihalcea(2000)]%
        {moldovan:2000}
\bibfield{author}{\bibinfo{person}{Dan~I. Moldovan} {and} \bibinfo{person}{Rada
  Mihalcea}.} \bibinfo{year}{2000}\natexlab{}.
\newblock \showarticletitle{Using WordNet and Lexical Operators to Improve
  Internet Searches}.
\newblock \bibinfo{journal}{\emph{IEEE Internet Computing}}
  \bibinfo{volume}{4}, \bibinfo{number}{1} (\bibinfo{date}{jan}
  \bibinfo{year}{2000}), \bibinfo{pages}{34–43}.
\newblock
\showISSN{1089-7801}
\urldef\tempurl%
\url{https://doi.org/10.1109/4236.815847}
\showDOI{\tempurl}


\bibitem[Morton et~al\mbox{.}(2014)]%
        {morton2014public}
\bibfield{author}{\bibinfo{person}{Kristi Morton}, \bibinfo{person}{Magdalena
  Balazinska}, \bibinfo{person}{Dan Grossman}, \bibinfo{person}{Robert Kosara},
  {and} \bibinfo{person}{Jock Mackinlay}.} \bibinfo{year}{2014}\natexlab{}.
\newblock \showarticletitle{Public Data and Visualizations: How are Many Eyes
  and Tableau Public used for Collaborative Analytics?}
\newblock \bibinfo{journal}{\emph{ACM SIGMOD Record}} \bibinfo{volume}{43},
  \bibinfo{number}{2} (\bibinfo{year}{2014}), \bibinfo{pages}{17--22}.
\newblock


\bibitem[Observable(2023a)]%
        {Observable}
\bibfield{author}{\bibinfo{person}{Observable}.}
  \bibinfo{year}{2023}\natexlab{a}.
\newblock \bibinfo{title}{Observable Notebooks}.
\newblock \bibinfo{howpublished}{\url{https://observablehq.com/}}.
\newblock


\bibitem[Observable(2023b)]%
        {Observable:searchtags}
\bibfield{author}{\bibinfo{person}{Observable}.}
  \bibinfo{year}{2023}\natexlab{b}.
\newblock \bibinfo{title}{Observable Notebooks}.
\newblock
  \bibinfo{howpublished}{\url{https://observablehq.com/@observablehq/searching-on-observable\#attributes}}.
\newblock


\bibitem[{OpenAI}(2023)]%
        {chatgpt}
\bibfield{author}{\bibinfo{person}{{OpenAI}}.} \bibinfo{year}{2023}\natexlab{}.
\newblock \bibinfo{title}{ChatGPT}.
\newblock
\newblock
\urldef\tempurl%
\url{https://openai.com/blog/chatgpt}
\showURL{%
Retrieved March 23, 2023 from \tempurl}


\bibitem[Oren et~al\mbox{.}(2008)]%
        {oren:2008}
\bibfield{author}{\bibinfo{person}{Eyal Oren}, \bibinfo{person}{Christophe
  Gu\'{e}ret}, {and} \bibinfo{person}{Stefan Schlobach}.}
  \bibinfo{year}{2008}\natexlab{}.
\newblock \showarticletitle{Anytime Query Answering in RDF through Evolutionary
  Algorithms}. In \bibinfo{booktitle}{\emph{Proceedings of the 7th
  International Conference on The Semantic Web}} (Karlsruhe, Germany)
  \emph{(\bibinfo{series}{ISWC '08})}. \bibinfo{publisher}{Springer-Verlag},
  \bibinfo{address}{Berlin, Heidelberg}, \bibinfo{pages}{98–113}.
\newblock
\showISBNx{9783540885634}
\urldef\tempurl%
\url{https://doi.org/10.1007/978-3-540-88564-1_7}
\showDOI{\tempurl}


\bibitem[Palani et~al\mbox{.}(2022)]%
        {interweave}
\bibfield{author}{\bibinfo{person}{Srishti Palani}, \bibinfo{person}{Yingyi
  Zhou}, \bibinfo{person}{Sheldon Zhu}, {and} \bibinfo{person}{Steven~P. Dow}.}
  \bibinfo{year}{2022}\natexlab{}.
\newblock \showarticletitle{InterWeave: Presenting Search Suggestions in
  Context Scaffolds Information Search and Synthesis}. In
  \bibinfo{booktitle}{\emph{Proceedings of the 35th Annual ACM Symposium on
  User Interface Software and Technology}} (Bend, OR, USA)
  \emph{(\bibinfo{series}{UIST '22})}. \bibinfo{publisher}{Association for
  Computing Machinery}, \bibinfo{address}{New York, NY, USA}, Article
  \bibinfo{articleno}{93}, \bibinfo{numpages}{16}~pages.
\newblock
\showISBNx{9781450393201}
\urldef\tempurl%
\url{https://doi.org/10.1145/3526113.3545696}
\showDOI{\tempurl}


\bibitem[{Programmable Web}(2022)]%
        {thesaurus}
\bibfield{author}{\bibinfo{person}{{Programmable Web}}.}
  \bibinfo{year}{2022}\natexlab{}.
\newblock \bibinfo{booktitle}{\emph{Thesaurus {API}}}.
\newblock
\urldef\tempurl%
\url{http://www.programmableweb.com/apitag/thesaurus}
\showURL{%
\tempurl}


\bibitem[Raman and Hellerstein(2001)]%
        {potterswheel}
\bibfield{author}{\bibinfo{person}{Vijayshankar Raman} {and}
  \bibinfo{person}{Joseph~M. Hellerstein}.} \bibinfo{year}{2001}\natexlab{}.
\newblock \showarticletitle{Potter's Wheel: An Interactive Data Cleaning
  System}. In \bibinfo{booktitle}{\emph{Proceedings of the 27th International
  Conference on Very Large Data Bases}} \emph{(\bibinfo{series}{VLDB '01})}.
  \bibinfo{publisher}{Morgan Kaufmann Publishers Inc.}, \bibinfo{address}{San
  Francisco, CA, USA}, \bibinfo{pages}{381–390}.
\newblock
\showISBNx{1558608044}


\bibitem[Ramos and Eickhoff(2020)]%
        {ramos:2020}
\bibfield{author}{\bibinfo{person}{Jerome Ramos} {and} \bibinfo{person}{Carsten
  Eickhoff}.} \bibinfo{year}{2020}\natexlab{}.
\newblock \showarticletitle{Search Result Explanations Improve Efficiency and
  Trust}. In \bibinfo{booktitle}{\emph{Proceedings of the 43rd International
  ACM SIGIR Conference on Research and Development in Information Retrieval}}
  (Virtual Event, China) \emph{(\bibinfo{series}{SIGIR '20})}.
  \bibinfo{publisher}{Association for Computing Machinery},
  \bibinfo{address}{New York, NY, USA}, \bibinfo{pages}{1597–1600}.
\newblock
\showISBNx{9781450380164}
\urldef\tempurl%
\url{https://doi.org/10.1145/3397271.3401279}
\showDOI{\tempurl}


\bibitem[Satyanarayan et~al\mbox{.}(2016)]%
        {satyanarayan2016vega}
\bibfield{author}{\bibinfo{person}{Arvind Satyanarayan},
  \bibinfo{person}{Dominik Moritz}, \bibinfo{person}{Kanit Wongsuphasawat},
  {and} \bibinfo{person}{Jeffrey Heer}.} \bibinfo{year}{2016}\natexlab{}.
\newblock \showarticletitle{Vega-lite: A grammar of interactive graphics}.
\newblock \bibinfo{journal}{\emph{IEEE Transactions on Visualization and
  Computer Graphics}} \bibinfo{volume}{23}, \bibinfo{number}{1}
  (\bibinfo{year}{2016}), \bibinfo{pages}{341--350}.
\newblock


\bibitem[Sechler et~al\mbox{.}(2017)]%
        {sechler2017sightline}
\bibfield{author}{\bibinfo{person}{Jordan Sechler}, \bibinfo{person}{Lane
  Harrison}, {and} \bibinfo{person}{Evan~M Peck}.}
  \bibinfo{year}{2017}\natexlab{}.
\newblock \showarticletitle{Sightline: Building on the web's visualization
  ecosystem}. In \bibinfo{booktitle}{\emph{Proceedings of the 2017 CHI Extended
  Abstracts}}. \bibinfo{pages}{2049--2055}.
\newblock


\bibitem[Setlur et~al\mbox{.}(2016)]%
        {eviza}
\bibfield{author}{\bibinfo{person}{Vidya Setlur}, \bibinfo{person}{Sarah~E.
  Battersby}, \bibinfo{person}{Melanie Tory}, \bibinfo{person}{Rich
  Gossweiler}, {and} \bibinfo{person}{Angel~X. Chang}.}
  \bibinfo{year}{2016}\natexlab{}.
\newblock \showarticletitle{Eviza: A Natural Language Interface for Visual
  Analysis}. In \bibinfo{booktitle}{\emph{Proceedings of the 29th Annual
  Symposium on User Interface Software and Technology}} (Tokyo, Japan)
  \emph{(\bibinfo{series}{UIST 2016})}. \bibinfo{publisher}{ACM},
  \bibinfo{address}{New York, NY, USA}, \bibinfo{pages}{365--377}.
\newblock
\showISBNx{978-1-4503-4189-9}


\bibitem[Setlur et~al\mbox{.}(2020)]%
        {sneakpique}
\bibfield{author}{\bibinfo{person}{Vidya Setlur}, \bibinfo{person}{Enamul
  Hoque}, \bibinfo{person}{Dae~Hyun Kim}, {and} \bibinfo{person}{Angel~X.
  Chang}.} \bibinfo{year}{2020}\natexlab{}.
\newblock \showarticletitle{Sneak Pique: Exploring Autocompletion as a Data
  Discovery Scaffold for Supporting Visual Analysis}. In
  \bibinfo{booktitle}{\emph{Proceedings of the 33rd Annual ACM Symposium on
  User Interface Software and Technology}} (Virtual Event, USA)
  \emph{(\bibinfo{series}{UIST '20})}. \bibinfo{publisher}{Association for
  Computing Machinery}, \bibinfo{address}{New York, NY, USA},
  \bibinfo{pages}{966–978}.
\newblock
\showISBNx{9781450375146}
\urldef\tempurl%
\url{https://doi.org/10.1145/3379337.3415813}
\showDOI{\tempurl}


\bibitem[Setlur et~al\mbox{.}(2019)]%
        {setlur2019inferencing}
\bibfield{author}{\bibinfo{person}{Vidya Setlur}, \bibinfo{person}{Melanie
  Tory}, {and} \bibinfo{person}{Alex Djalali}.}
  \bibinfo{year}{2019}\natexlab{}.
\newblock \showarticletitle{Inferencing Underspecified Natural Language
  Utterances in Visual Analysis}. In \bibinfo{booktitle}{\emph{Proceedings of
  the 24th International Conference on Intelligent User Interfaces}} (Marina
  del Ray, California) \emph{(\bibinfo{series}{IUI '19})}.
  \bibinfo{publisher}{Association for Computing Machinery},
  \bibinfo{address}{New York, NY, USA}, \bibinfo{pages}{40--51}.
\newblock
\showISBNx{9781450362726}
\urldef\tempurl%
\url{https://doi.org/10.1145/3301275.3302270}
\showDOI{\tempurl}


\bibitem[Shen et~al\mbox{.}(2021)]%
        {shen2021towards}
\bibfield{author}{\bibinfo{person}{Leixian Shen}, \bibinfo{person}{Enya Shen},
  \bibinfo{person}{Yuyu Luo}, \bibinfo{person}{Xiaocong Yang},
  \bibinfo{person}{Xuming Hu}, \bibinfo{person}{Xiongshuai Zhang},
  \bibinfo{person}{Zhiwei Tai}, {and} \bibinfo{person}{Jianmin Wang}.}
  \bibinfo{year}{2021}\natexlab{}.
\newblock \showarticletitle{Towards natural language interfaces for data
  visualization: A survey}.
\newblock \bibinfo{journal}{\emph{arXiv preprint arXiv:2109.03506}}
  (\bibinfo{year}{2021}).
\newblock


\bibitem[Shokouhi and Si(2011)]%
        {Shokouhi2011FederatedS}
\bibfield{author}{\bibinfo{person}{Milad Shokouhi} {and} \bibinfo{person}{Luo
  Si}.} \bibinfo{year}{2011}\natexlab{}.
\newblock \showarticletitle{Federated Search}. In
  \bibinfo{booktitle}{\emph{Foundations and Trends in Information Retrieval}}.
\newblock


\bibitem[Silverstein et~al\mbox{.}(1999)]%
        {silverstein1999analysis}
\bibfield{author}{\bibinfo{person}{Craig Silverstein}, \bibinfo{person}{Hannes
  Marais}, \bibinfo{person}{Monika Henzinger}, {and} \bibinfo{person}{Michael
  Moricz}.} \bibinfo{year}{1999}\natexlab{}.
\newblock \showarticletitle{Analysis of a very large web search engine query
  log}. In \bibinfo{booktitle}{\emph{Acm sigir forum}},
  Vol.~\bibinfo{volume}{33}. ACM New York, NY, USA, \bibinfo{pages}{6--12}.
\newblock


\bibitem[Sivic and Zisserman(2003)]%
        {videogoogle}
\bibfield{author}{\bibinfo{person}{Sivic} {and} \bibinfo{person}{Zisserman}.}
  \bibinfo{year}{2003}\natexlab{}.
\newblock \showarticletitle{Video Google: a text retrieval approach to object
  matching in videos}. In \bibinfo{booktitle}{\emph{Proceedings Ninth IEEE
  International Conference on Computer Vision}}. \bibinfo{pages}{1470--1477
  vol.2}.
\newblock
\urldef\tempurl%
\url{https://doi.org/10.1109/ICCV.2003.1238663}
\showDOI{\tempurl}


\bibitem[Smeulders et~al\mbox{.}(2000)]%
        {cbr:2000}
\bibfield{author}{\bibinfo{person}{A.W.M. Smeulders}, \bibinfo{person}{M.
  Worring}, \bibinfo{person}{S. Santini}, \bibinfo{person}{A. Gupta}, {and}
  \bibinfo{person}{R. Jain}.} \bibinfo{year}{2000}\natexlab{}.
\newblock \showarticletitle{Content-based image retrieval at the end of the
  early years}.
\newblock \bibinfo{journal}{\emph{IEEE Transactions on Pattern Analysis and
  Machine Intelligence}} \bibinfo{volume}{22}, \bibinfo{number}{12}
  (\bibinfo{year}{2000}), \bibinfo{pages}{1349--1380}.
\newblock
\urldef\tempurl%
\url{https://doi.org/10.1109/34.895972}
\showDOI{\tempurl}


\bibitem[Srihari and Li(1999)]%
        {Srihari1999InformationES}
\bibfield{author}{\bibinfo{person}{Rohini~K. Srihari} {and} \bibinfo{person}{W.
  Li}.} \bibinfo{year}{1999}\natexlab{}.
\newblock \showarticletitle{Information Extraction Supported Question
  Answering}. In \bibinfo{booktitle}{\emph{Text Retrieval Conference}}.
\newblock


\bibitem[Srinivasan et~al\mbox{.}(2018)]%
        {srinivasan2018augmenting}
\bibfield{author}{\bibinfo{person}{Arjun Srinivasan}, \bibinfo{person}{Steven~M
  Drucker}, \bibinfo{person}{Alex Endert}, {and} \bibinfo{person}{John
  Stasko}.} \bibinfo{year}{2018}\natexlab{}.
\newblock \showarticletitle{Augmenting visualizations with interactive data
  facts to facilitate interpretation and communication}.
\newblock \bibinfo{journal}{\emph{IEEE transactions on visualization and
  computer graphics}} \bibinfo{volume}{25}, \bibinfo{number}{1}
  (\bibinfo{year}{2018}), \bibinfo{pages}{672--681}.
\newblock


\bibitem[Srinivasan et~al\mbox{.}(2021)]%
        {srinivasan2021collecting}
\bibfield{author}{\bibinfo{person}{Arjun Srinivasan}, \bibinfo{person}{Nikhila
  Nyapathy}, \bibinfo{person}{Bongshin Lee}, \bibinfo{person}{Steven~M
  Drucker}, {and} \bibinfo{person}{John Stasko}.}
  \bibinfo{year}{2021}\natexlab{}.
\newblock \showarticletitle{Collecting and characterizing natural language
  utterances for specifying data visualizations}. In
  \bibinfo{booktitle}{\emph{Proceedings of the 2021 CHI Conference on Human
  Factors in Computing Systems}}. \bibinfo{pages}{1--10}.
\newblock


\bibitem[Srinivasan and Setlur(2021)]%
        {srinivasan2021snowy}
\bibfield{author}{\bibinfo{person}{Arjun Srinivasan} {and}
  \bibinfo{person}{Vidya Setlur}.} \bibinfo{year}{2021}\natexlab{}.
\newblock \showarticletitle{Snowy: Recommending utterances for conversational
  visual analysis}. In \bibinfo{booktitle}{\emph{The 34th Annual ACM Symposium
  on User Interface Software and Technology}}. \bibinfo{pages}{864--880}.
\newblock


\bibitem[Srinivasan and Stasko(2018)]%
        {orko}
\bibfield{author}{\bibinfo{person}{Arjun Srinivasan} {and}
  \bibinfo{person}{John Stasko}.} \bibinfo{year}{2018}\natexlab{}.
\newblock \showarticletitle{Orko: Facilitating multimodal interaction for
  visual exploration and analysis of networks}.
\newblock \bibinfo{journal}{\emph{IEEE Transactions on Visualization and
  Computer Graphics}} \bibinfo{volume}{24}, \bibinfo{number}{1}
  (\bibinfo{year}{2018}), \bibinfo{pages}{511--521}.
\newblock


\bibitem[{Tableau Software}(2023)]%
        {tableau2023public}
\bibfield{author}{\bibinfo{person}{{Tableau Software}}.}
  \bibinfo{year}{2023}\natexlab{}.
\newblock \bibinfo{title}{Tableau {P}ublic}.
\newblock
\newblock
\urldef\tempurl%
\url{https://public.tableau.com/}
\showURL{%
Retrieved March 23, 2023 from \tempurl}


\bibitem[Thomas et~al\mbox{.}(2007)]%
        {Thomas2007ONTOSEARCH2SO}
\bibfield{author}{\bibinfo{person}{Edward Thomas}, \bibinfo{person}{Jeff~Z.
  Pan}, {and} \bibinfo{person}{Derek~H. Sleeman}.}
  \bibinfo{year}{2007}\natexlab{}.
\newblock \showarticletitle{ONTOSEARCH2: Searching Ontologies Semantically}. In
  \bibinfo{booktitle}{\emph{OWL: Experiences and Directions}}.
\newblock


\bibitem[Tory and Setlur(2019)]%
        {tory2019mean}
\bibfield{author}{\bibinfo{person}{Melanie Tory} {and} \bibinfo{person}{Vidya
  Setlur}.} \bibinfo{year}{2019}\natexlab{}.
\newblock \showarticletitle{Do what i mean, not what i say! design
  considerations for supporting intent and context in analytical conversation}.
  In \bibinfo{booktitle}{\emph{2019 IEEE conference on visual analytics science
  and technology (VAST)}}. IEEE, \bibinfo{pages}{93--103}.
\newblock


\bibitem[Tran et~al\mbox{.}(2007)]%
        {Tran2007OntologyBasedIO}
\bibfield{author}{\bibinfo{person}{Thanh Tran}, \bibinfo{person}{Philipp
  Cimiano}, \bibinfo{person}{Sebastian Rudolph}, {and} \bibinfo{person}{Rudi
  Studer}.} \bibinfo{year}{2007}\natexlab{}.
\newblock \showarticletitle{Ontology-Based Interpretation of Keywords for
  Semantic Search}. In \bibinfo{booktitle}{\emph{ISWC/ASWC}}.
\newblock


\bibitem[Vaca-Castano and Shah(2015)]%
        {vaca-castano:2015}
\bibfield{author}{\bibinfo{person}{Gonzalo Vaca-Castano} {and}
  \bibinfo{person}{Mubarak Shah}.} \bibinfo{year}{2015}\natexlab{}.
\newblock \showarticletitle{Semantic Image Search From Multiple Query Images}.
  In \bibinfo{booktitle}{\emph{Proceedings of the 23rd ACM International
  Conference on Multimedia}} (Brisbane, Australia) \emph{(\bibinfo{series}{MM
  '15})}. \bibinfo{publisher}{Association for Computing Machinery},
  \bibinfo{address}{New York, NY, USA}, \bibinfo{pages}{887–890}.
\newblock
\showISBNx{9781450334594}
\urldef\tempurl%
\url{https://doi.org/10.1145/2733373.2806356}
\showDOI{\tempurl}


\bibitem[Viegas et~al\mbox{.}(2007)]%
        {viegas2007manyeyes}
\bibfield{author}{\bibinfo{person}{Fernanda~B Viegas}, \bibinfo{person}{Martin
  Wattenberg}, \bibinfo{person}{Frank Van~Ham}, \bibinfo{person}{Jesse Kriss},
  {and} \bibinfo{person}{Matt McKeon}.} \bibinfo{year}{2007}\natexlab{}.
\newblock \showarticletitle{Manyeyes: a site for visualization at internet
  scale}.
\newblock \bibinfo{journal}{\emph{IEEE Transactions on Visualization and
  Computer Graphics}} \bibinfo{volume}{13}, \bibinfo{number}{6}
  (\bibinfo{year}{2007}), \bibinfo{pages}{1121--1128}.
\newblock


\bibitem[Wang et~al\mbox{.}(2007)]%
        {panto}
\bibfield{author}{\bibinfo{person}{Chong Wang}, \bibinfo{person}{Miao Xiong},
  \bibinfo{person}{Qi Zhou}, {and} \bibinfo{person}{Yong Yu}.}
  \bibinfo{year}{2007}\natexlab{}.
\newblock \showarticletitle{PANTO: A Portable Natural Language Interface to
  Ontologies}. In \bibinfo{booktitle}{\emph{Proceedings of the 4th European
  Conference on The Semantic Web: Research and Applications}} (Innsbruck,
  Austria) \emph{(\bibinfo{series}{ESWC '07})}.
  \bibinfo{publisher}{Springer-Verlag}, \bibinfo{address}{Berlin, Heidelberg},
  \bibinfo{pages}{473–487}.
\newblock
\showISBNx{9783540726661}
\urldef\tempurl%
\url{https://doi.org/10.1007/978-3-540-72667-8_34}
\showDOI{\tempurl}


\bibitem[Willett et~al\mbox{.}(2007)]%
        {willett2007scented}
\bibfield{author}{\bibinfo{person}{Wesley Willett}, \bibinfo{person}{Jeffrey
  Heer}, {and} \bibinfo{person}{Maneesh Agrawala}.}
  \bibinfo{year}{2007}\natexlab{}.
\newblock \showarticletitle{Scented widgets: Improving navigation cues with
  embedded visualizations}.
\newblock \bibinfo{journal}{\emph{IEEE transactions on visualization and
  computer graphics}} \bibinfo{volume}{13}, \bibinfo{number}{6}
  (\bibinfo{year}{2007}), \bibinfo{pages}{1129--1136}.
\newblock


\bibitem[Wu and Palmer(1994)]%
        {wu1994verbs}
\bibfield{author}{\bibinfo{person}{Zhibiao Wu} {and} \bibinfo{person}{Martha
  Palmer}.} \bibinfo{year}{1994}\natexlab{}.
\newblock \showarticletitle{Verbs semantics and lexical selection}. In
  \bibinfo{booktitle}{\emph{Proceedings of ACL}}. \bibinfo{pages}{133--138}.
\newblock


\bibitem[Younger(1967)]%
        {younger1967recognition}
\bibfield{author}{\bibinfo{person}{Daniel~H Younger}.}
  \bibinfo{year}{1967}\natexlab{}.
\newblock \showarticletitle{Recognition and parsing of context-free languages
  in time n3}.
\newblock \bibinfo{journal}{\emph{Information and control}}
  \bibinfo{volume}{10}, \bibinfo{number}{2} (\bibinfo{year}{1967}),
  \bibinfo{pages}{189--208}.
\newblock


\bibitem[Yujian and Bo(2007)]%
        {yujian2007normalized}
\bibfield{author}{\bibinfo{person}{Li Yujian} {and} \bibinfo{person}{Liu Bo}.}
  \bibinfo{year}{2007}\natexlab{}.
\newblock \showarticletitle{A normalized Levenshtein distance metric}.
\newblock \bibinfo{journal}{\emph{IEEE Transactions on Pattern Analysis and
  Machine Intelligence}} \bibinfo{volume}{29}, \bibinfo{number}{6}
  (\bibinfo{year}{2007}), \bibinfo{pages}{1091--1095}.
\newblock


\bibitem[Zenz et~al\mbox{.}(2009)]%
        {zenz:2009}
\bibfield{author}{\bibinfo{person}{Gideon Zenz}, \bibinfo{person}{Xuan Zhou},
  \bibinfo{person}{Enrico Minack}, \bibinfo{person}{Wolf Siberski}, {and}
  \bibinfo{person}{Wolfgang Nejdl}.} \bibinfo{year}{2009}\natexlab{}.
\newblock \showarticletitle{From Keywords to Semantic Queries-Incremental Query
  Construction on the Semantic Web}.
\newblock \bibinfo{journal}{\emph{Web Semant.}} \bibinfo{volume}{7},
  \bibinfo{number}{3} (\bibinfo{date}{sep} \bibinfo{year}{2009}),
  \bibinfo{pages}{166–176}.
\newblock
\showISSN{1570-8268}
\urldef\tempurl%
\url{https://doi.org/10.1016/j.websem.2009.07.005}
\showDOI{\tempurl}


\bibitem[Zhai et~al\mbox{.}(2019)]%
        {visualsearchpinterest}
\bibfield{author}{\bibinfo{person}{Andrew Zhai}, \bibinfo{person}{Hao-Yu Wu},
  \bibinfo{person}{Eric Tzeng}, \bibinfo{person}{Dong~Huk Park}, {and}
  \bibinfo{person}{Charles Rosenberg}.} \bibinfo{year}{2019}\natexlab{}.
\newblock \showarticletitle{Learning a Unified Embedding for Visual Search at
  Pinterest}. In \bibinfo{booktitle}{\emph{Proceedings of the 25th ACM SIGKDD
  International Conference on Knowledge Discovery \& Data Mining}} (Anchorage,
  AK, USA) \emph{(\bibinfo{series}{KDD '19})}. \bibinfo{publisher}{Association
  for Computing Machinery}, \bibinfo{address}{New York, NY, USA},
  \bibinfo{pages}{2412–2420}.
\newblock
\showISBNx{9781450362016}
\urldef\tempurl%
\url{https://doi.org/10.1145/3292500.3330739}
\showDOI{\tempurl}


\end{thebibliography}

\end{document}